%% file: perfect_bipartite_matching.tex
\documentclass[10pt,twoside,a4paper]{article}

\usepackage{a4wide}
\usepackage{amssymb,amsmath,amsthm}

\usepackage[algosection,algoruled,vlined,nofillcomment]{algorithm2e}
\SetEndCharOfAlgoLine{}

\usepackage[colorlinks=true,citecolor=blue,urlcolor=blue,linkcolor=blue,bookmarksopen=true,unicode=true,pdffitwindow=true]{hyperref}

\usepackage{tikz}
\usetikzlibrary{positioning}

\renewcommand{\O}[1]{\ensuremath{\mathcal{O}{\left( #1 \right)}}}

\newcommand{\set}[1]{\ensuremath{\left\{#1 \right\}}}

\newcommand{\M}{\mathcal{M}}
\newcommand{\MM}{\hat{\M}}

\newcommand{\K}{\mathcal{K}}
\newcommand{\hG}{\hat{G_\pi}}
\newcommand{\Ex}{E_{\text{ex}}}

\newcommand{\dotcup}{\dot{\cup}}

\DeclareMathOperator*{\Square}{\square}

\DeclareMathOperator{\scc}{cc}

\DeclareMathOperator{\up}{up}
\DeclareMathOperator{\visit}{visit}
\DeclareMathOperator{\xskip}{skip}
\DeclareMathOperator{\xleft}{left}
\DeclareMathOperator{\xright}{right}

\newtheorem{theorem}{Theorem}[section]
\newtheorem{observation}[theorem]{Observation}
\newtheorem{lemma}[theorem]{Lemma}

\SetKwFunction{Enumerate}{Enumerate}
\SetKwFunction{Visit}{Visit}
\SetKwFunction{Ready}{MakeReady}
\SetKwFunction{Trim}{Trim}
\SetKwFunction{Main}{Main}
\SetKwProg{myproc}{Procedure}{}{}

\title{Constant time enumeration of perfect bipartite matchings}
\author{Ji\v{r}\'i Fink \footnote{E-mail: \href{mailto:fink@ktiml.mff.cuni.cz}{fink@ktiml.mff.cuni.cz}} \\ \small Department of Theoretical Computer Science and Mathematical Logic\\\small Faculty of Mathematics and Physics\\\small Charles University in Prague}
\date{}

\begin{document}
\maketitle
\begin{abstract}
We present an algorithm that enumerates all the perfect matchings in a given bipartite graph $G = (V,E)$.
Our algorithm requires a constant amortized time to visit one perfect matching of $G$, in contrast to the current fastest algorithm, published 25 years ago by Uno~\cite{uno2001fast}, which requires $\O{\log |V|}$ time.

To facilitate the listing of all edges in a visited perfect matching, we develop a variant of arithmetic circuits, which may have broader applications in future enumeration algorithms.
Consequently, a visited perfect matching is represented within a binary tree.
Although it is more common to provide visited objects in an array, we present a class of graphs for which achieving constant amortized time is not feasible in this case.
\end{abstract}

\section{Introduction}

Given a graph $G = (V,E)$, a \emph{perfect matching} is a subset $M \subseteq E$ such that every vertex in $V$ is incident to exactly one edge in $M$.
Perfect matchings in bipartite graphs are classical combinatorial structures with extensive applications in optimization, economics, scheduling, chemistry, and computer science; see e.g.~\cite{babic2002combinatorial,marino2015analysis,schrijver1998theory,gale1962college}.

One of the foundational results in graph theory is characterizing bipartite graphs that has a perfect matching.
The famous \emph{Hall's marriage theorem}~\cite{hall1935representatives} states that a bipartite graph $G$ has a perfect matching if and only if, for every subset $W$ of vertices of one pertite satisfies $|N_V(W)| \ge |W|$, where $N_V(W)$ is the set of neighbors of $W$.

Perfect matchings also play a crucial role in polyhedral combinatorics.
The \emph{perfect matching polytope} -- defined as the convex hull of incidence vectors of all perfect matchings in a graph -- has been extensively studied and has influenced the development of combinatorial optimization theory; see Edmonds~\cite{edmonds1965paths}, Schrijver~\cite{schrijver1998theory}, and Lovász and Plummer~\cite{lovasz2009matching}.

\medskip
\textbf{Optimization and Counting.}
The \emph{assignment problem} is one of the most important problems in combinatorial optimization.
Given a bipartite graph with edge costs, the task is to find a perfect matching of minimum total cost.
The classical Hungarian algorithm~\cite{kuhn1955hungarian} solves this problem in polynomial time; see also~\cite{pentico2007assignment}.
The Hopcroft-Karp algorithm~\cite{hopcroft1973n} finds a maximum matching in bipartite graphs in $\mathcal{O}(|E| \sqrt{|V|})$ time.
More recently, Chen, Kyng, Liu, Peng~\cite{chen2022maximum} presented a breakthrough algorithm that solves the same problem in $\mathcal{O}(|E|^{1+o(1)})$ time.

In stark contrast to optimization, the problem of \emph{counting} perfect matchings is much harder.
Valiant~\cite{valiant1979complexity} showed that counting the number of perfect matchings in a bipartite graph, which is equivalent to computing the permanent of a $0/1$ matrix, is \#P-complete.

\medskip
\textbf{Enumeration.}
Given the high complexity of counting, a natural question arises: can we efficiently \emph{enumerate} all perfect matchings?
Enumeration algorithms aim to list all objects in a given class, and their efficiency is usually measured in terms of \emph{delay} -- the time required to output the next objects after the previous one.

The ideal scenario is to achieve \emph{constant delay}, also called \emph{loopless enumeration}~\cite{ehrlich1973loopless,knuth2011art,mutze2022combinatorial}.
Loopless algorithms are known for simple combinatorial objects such as permutations, binary strings, and bounded subsets~\cite{ehrlich1973loopless}, and also for binary trees~\cite{ruskey2008generating}.
For more advanced objects, it is reasonable to relax this worst-case measure, and to consider the amortized delay, i.e., the total time spent to generate a class of objects $X$, divided by the cardinality of $X$. 
In such cases, it is common to aim for \emph{constant amortized delay} (CAT). Constant amortized time enumeration algorithms have been developed for several combinatorial, e.g. spanning trees~\cite{kapoor1991algorithms}, feasible solutions to knapsack problems~\cite{sawada2012efficient}, matchings~\cite{uno2015constant} and connected induced subgraphs~\cite{uno2015constant}.
An extensive survey of such results is provided in~\cite{mutze2022combinatorial}.

Pruesse and Ruskey~\cite{pruesse1994generating} posed a fundamental question: does every \#P-complete problem admit a CAT enumeration algorithm?
They gave a positive answer for the case of enumerating all linear extensions of a poset, a problem known to be \#P-complete~\cite{brightwell1991counting}.
Later improvements by Canfield and Williamson~\cite{canfield1995loop} even yielded a loopless version.
Kurita and Wasa~\cite{kurita2022constant} developed a CAT algorithm for Eulerian trails, and Brightwell and Winkler~\cite{brightwell2005counting} showed that counting such trails is also \#P-complete.

\medskip
\textbf{Enumeration of Perfect Matchings.}
The problem of enumerating all perfect matchings in a bipartite graph $G$ has a long history.
The first such algorithm was proposed by Itai, Rodeh, and Tanimoto~\cite{tanimoto1978some} in 1978, with amortized delay $\mathcal{O}(|E|)$.
Independently, the same delay was achieved by Fukuda and Matsui~\cite{fukuda1994finding}.
In 1997, Uno~\cite{uno1997algorithms} improved the delay to $\mathcal{O}(|V|)$, and in 2001, presented the first sublinear-delay algorithm with amortized delay $\mathcal{O}(\log |V|)$~\cite{uno2001fast}.
However, this last algorithm only outputs a fixed string ``matching'' instead of the list of edges of the visited perfect matching.

\medskip
\textbf{Our Contributions.}
In this paper, we present a new enumeration algorithm that visits the set of all perfect matchings, denoted by $\M(G)$, shortly $\M$, of a bipartite graph $G$ in constant amortized time while also providing a binary tree that contains all the edges of the visited perfect matchings.

\begin{theorem}\label{thm:main}
All perfect matchings $\M$ of a given bipartite graph $G = (V,E)$ can be enumerated in total time $\O{|\M| + |V|^2 + |E|^{1+o(1)}}$ and space $\mathcal{O}(|V|^2)$.
\end{theorem}

The total time consists of: $\mathcal{O}(|E|^{1+o(1)})$ to find the first perfect matching~\cite{chen2022maximum}, $\mathcal{O}(|V|^2)$ for initialization, and constant amortized time for generating all perfect matchings.

\begin{figure}[ht]
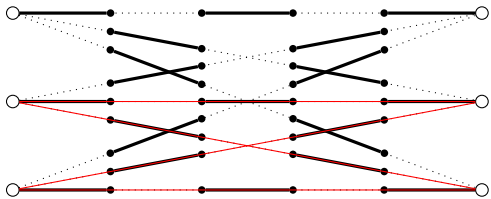

\centering
\include{array}
\caption{Graph $H_{3,5}$.
Empty vertices belong to $K_{n,n}$; small filled vertices appear on paths replacing edges.
Bold edges form one perfect matching; dotted edges are non-matching.
The red cycle is a minimal alternating cycle.}
\label{fig:array}
\end{figure}

As customary in enumeration literature~\cite{ruskey}, we exclude from the time complexity the cost of outputting or processing each object.
Although enumerated objects are usually provided by a pointer to an array storing the visited object, using the array is not possible in our case as the following example explains; see Figure \ref{fig:array}.

Consider the graph $H_{n,k}$ constructed by replacing each edge of the complete bipartite graph $K_{n,n}$ with a path on $k$ edges, for odd $k$.
The graph has $n!$ perfect matchings (inherited from $K_{n,n}$), but any two perfect matchings differ in at least $4k$ edges.
Hence, updating an output array to reflect a new matching requires $\Omega(k)$ time, which implies a total time of $\Omega(kn!)$.
However, our goal is achieving time complexity $\O{n^4 k^2 + n!}$ for the graph $H_{n,k}$.

To overcome this, our algorithm stores each enumerated perfect matching in a \emph{binary tree}, called the \emph{visiting tree}.
Each leaf corresponds to an edge in the matching, and each internal node has two children to ensure that all leaves (edges) can be listed in $\O{|V|}$ time.
If needed, leaves can be augmented with pointers to form a linked list, but we omit these technical details for clarity.

\medskip
\textbf{Algorithmic Techniques.}
Our algorithm is based on several known and novel ideas.
    
A classical technique in enumeration algorithm is known the flashlight search or binary partition method \cite{read1975bounds,bussieck1998vertex}.
Our recursive function obtains a graph $G$ and constructs graphs $G_A$ and $G_B$ such that both contains at least one perfect matching and $\M(G) = \M(G_A) \dotcup \M(G_B)$.

As presented in Figure \ref{fig:array}, vertices of degree two can increase the size of a graph while preserving the number of perfect matchings which may increase the delay.
Therefore, Uno \cite{uno2001fast} presented an operation called trimming which reduces the graph by removing pairs of neighbor vertices of degree two and modifying their incident edges.
We enhance this idea by removing all vertices of degree two and contracting their neighbor vertices.

Both versions of trimming do not change the number of perfect matchings but perfect matchings in the trimmed graph miss some edges of the original graph.
Therefore, our algorithm keeps track of the removed edges using a variant of arithmetic circuit.
An inspiration of this approach comes from Behle and Eisenbrand \cite{behle20070} where a binary decision diagram is constructed for given 0/1-polytope to efficiently enumerate of all its vertices.
Although their algorithm performs well in practice, it requires exponential space in the worst case.
Therefore, we combine both recursion and constructing our special circuit to achieve constant amortized delay and only $\O{|V|^2}$ space.

\medskip
\textbf{Extensions and Related Problems.}
Our framework also supports enumeration of all \emph{minimum-cost} perfect matchings in a bipartite graph.
In this case, use the dual LP solution and complementary slackness to extract a subset $F \subseteq E$ such that every subset $M$ of edges $E$ is a perfect matching in the subgraph $(V,F)$ if and only if $M$ is a minimum-cost perfect matching in $G$; see e.g. \cite{schrijver1998theory}.
Then apply our enumeration algorithm on the subgraph $(V, F)$.

A natural question is whether our approach can be generalized to other classes of matchings.
Tanimoto, Itai, Rodeh~\cite{tanimoto1978some} and Uno~\cite{uno1997algorithms} gave algorithms for enumerating all perfect, maximum, and maximal matchings in bipartite graphs with delay $\mathcal{O}(|E|)$ and $\mathcal{O}(|V|)$, respectively.
For general graphs, Gabow, Kaplan, Tarjan~\cite{gabow1999unique} gave an algorithm that tests whether a given perfect matching is unique in $\mathcal{O}(|E|)$ time, and similar results exist for maximum matchings.
These tests can be embedded in enumeration frameworks to achieve delay $\mathcal{O}(|E|)$, but whether constant amortized delay is achievable in general remains open.

There are many combinatorial problems related to perfect matchings in bipartite graphs.
For example, economics and game theory studies stable matchings \cite{gale1962college}.
Gusfield \cite{gusfield1987three} enumerates all stable matchings with delay $\O{n}$ and Ruskey \cite{ruskey} asked to develop a constant amortized time algorithm.

\medskip

The next section presents top-down overview of ideas used in the algorithm and following sections gives all details in the bottom-up approach.
A list of symbols used in this paper is provided in the appendix.


\section{Overview of our algorithm}

We begin by reviewing established results on perfect matchings in bipartite graphs and alternating cycles; see, for example, \cite{uno2001fast, uno1997algorithms, gabow1999unique, tanimoto1978some}.
Throughout this paper, we focus exclusively on bipartite graphs that contain at least one perfect matching.
Moreover, we assume that each graph is provided along with one of its perfect matchings.

Similar to the approach in \cite{uno2001fast}, our enumeration algorithm combines trimming and recursion.
Trimming eliminates certain vertices and edges that are not necessary for the enumeration, without altering the set of perfect matchings.
The recursive procedure \Enumerate operates on a graph $G$ by constructing two subgraphs, $G_A$ and $G_B$, such that $\M(G) = \M(G_A) \dotcup \M(G_B)$ where $\dotcup$ denotes a disjoint union.
The function then recursively enumerates matchings in both $G_A$ and $G_B$.

Let $N_V(u)$ denote the set of all neighbors of a vertex $u$ in $G$, and let $N_E(u)$ denote the set of all edges incident to $u$ in $G$.
For a set of vertices $W \subseteq V(G)$, we define $N_V(W) = \bigcup_{u \in W} N_V(u)$ and $N_E(W) = \bigcup_{u \in W} N_E(u)$.

\medskip
\textbf{Trimming edges.}

Trimming removes all edges of the graph $G$ that are not contained in any perfect matching.
Furthermore, edges contained in all perfect matchings are replaced by a single isolated edge.
This process can be implemented using the following approach.

Let $b^+(G)$ denote the set of edges contained in every perfect matching of $G$, and let $b^-(G)$ denote the set of edges contained in no perfect matching of $G$.
Define $b(G)$ as the union $b(G) = b^+(G) \cup b^-(G)$.
Let $D(G, M)$ be a directed graph constructed from a graph $G$ and a perfect matching $M \in \M$ as follows: each edge in $M$ is oriented from one partite set of $G$ to the other, while each edge in $E \setminus M$ is oriented in the opposite direction.

A cycle in $G$ is called \emph{$M$-alternating} if its edges alternate between those in $M$ and those not in $M$.
Note that $M$-alternating cycles in $G$ correspond to directed cycles in $D(G, M)$.

Given a perfect matching $M$ and an edge $e$, an $M$-alternating cycle containing $e$ can be found in time $\mathcal{O}(|E|)$ using alternating trees from the Hopcroft–Karp algorithm~\cite{hopcroft1973n}.
We frequently employ this algorithm to either find a directed cycle in $D(G, M)$ that includes a given edge, or to compute a perfect matching $M'$ such that $e \in M \triangle M'$, where $\triangle$ denotes the symmetric difference between two sets.
Both operations can be performed in time $\mathcal{O}(|E|)$.

The following lemma combines the characterization of edges in $b(G)$, due to \cite{tanimoto1978some}, with Tarjan's algorithm~\cite{tarjan1972depth} for identifying all bridges and strongly connected components in a directed graph in time $\mathcal{O}(|E|)$.

\begin{lemma} \label{lem:basic_structure}
The following statements are equivalent for every edge $e$ in a bipartite graph $G$ with a perfect matching $M$:
\begin{itemize}
    \item $e \in b(G)$;
    \item $e$ is a bridge in $D(G, M)$;
    \item there exists no $M$-alternating cycle in $G$ that contains $e$.
\end{itemize}
Moreover, the sets $b^+(G)$ and $b^-(G)$ can both be computed in time $\mathcal{O}(|E|)$.
The sets of bridges and the strongly connected components of $D(G, M)$ are identical for all perfect matchings $M \in \M$.
\end{lemma}

Let $G^b$ be the graph $G$ without edges $b^-(G)$.
Observe that an edge $uv$ belongs to $b^+(G)$ if and only if all other edges incident to either $u$ or $v$ belong to $b^-(G)$.
The edges in $b^+(G)$ form connected components within $G^b$, which we refer to as \emph{trivial components}.
The components of $G^b$ comprise the strongly connected components of $D(G, M)$ along with the trivial components.
We say that a bipartite graph $G$ is \emph{strongly connected} if the digraph $D(G, M)$ is strongly connected.

The first step in trimming the graph $G$ involves removing all edges in $b^-(G)$ and replacing all edges in $b^+(G)$ by a single isolated edge.
All edges in $b^+(G)$ are stored in a dedicated data structure, referred to as a \emph{union-product circuit}.
We first present the motivation for introducing this circuit, followed by its formal definition.

Finally, note that after handling the edges in $b(G)$, each vertex in the graph has degree at least two, except for the endpoints of at most one isolated edge.

\medskip
\textbf{Trimming vertices of degree 2.}

Let $u$ be a vertex of degree two with neighbors $v$ and $w$.
We perform trimming on $u$ as follows.

Let $v_1, \ldots, v_\ell$ denote all neighbors of $v$ excluding $u$, and let $w_1, \ldots, w_k$ denote all neighbors of $w$ excluding $u$.
It is straightforward to observe that the edge $uv$ is included in a perfect matching $M \in \M$ if and only if $M$ contains one of the edges $ww_1, \ldots, ww_k$.
Similarly, $uw \in M$ if and only if $M$ contains one of the edges $vv_1, \ldots, vv_\ell$.
Furthermore, $M$ contains exactly one edge of $ww_1, \ldots, ww_k, vv_1, \ldots, vv_\ell$

To capture this relationship, we construct a new graph $G'$ from $G$ by removing vertex $u$ and its incident edges, and contracting $v$ and $w$ into a new vertex $z$.
The vertex $z$ then connects to $v_1, \ldots, v_\ell, w_1, \ldots, w_k$.

Although graphs $G$ and $G'$ have the same number of perfect matchings, care must be taken when interpreting the edges in perfect matchings.
Specifically, for every edge $zw_i$ included in a perfect matching of $G'$, we must instead output edges $uv$ and $ww_i$ in the original graph $G$.
Similarly, if $zv_i$ is in a perfect matching of $G'$, it corresponds to edges $uw$ and $vv_i$ in $G$.
To support this transformation, each edge in $G'$ is associated with a node in a \emph{union-product circuit}, which allows the correct expansion of each edge back into the original graph.

When $v$ and $w$ share a common neighbor other than $u$, the contraction may result in parallel edges in $G'$, say $e_1$ and $e_2$.
For every perfect matching in $G'$ containing $e_1$, there exists another perfect matching that differs only by replacing $e_1$ with $e_2$.
In such cases, we can safely remove $e_2$, but we must record that two matchings correspond to $e_1$, and both must be considered during enumeration.

Moreover, $G'$ may contain other vertices with degree 2, leading to further contractions and additional parallel edges, thereby multiplying the number of perfect matchings.
We emphasize that parallel edges are only introduced as an intermediate step during the trimming of a single vertex of degree two.
Once the trimming step is completed, no parallel edges remain in the resulting graph.

\begin{figure}[tbp]
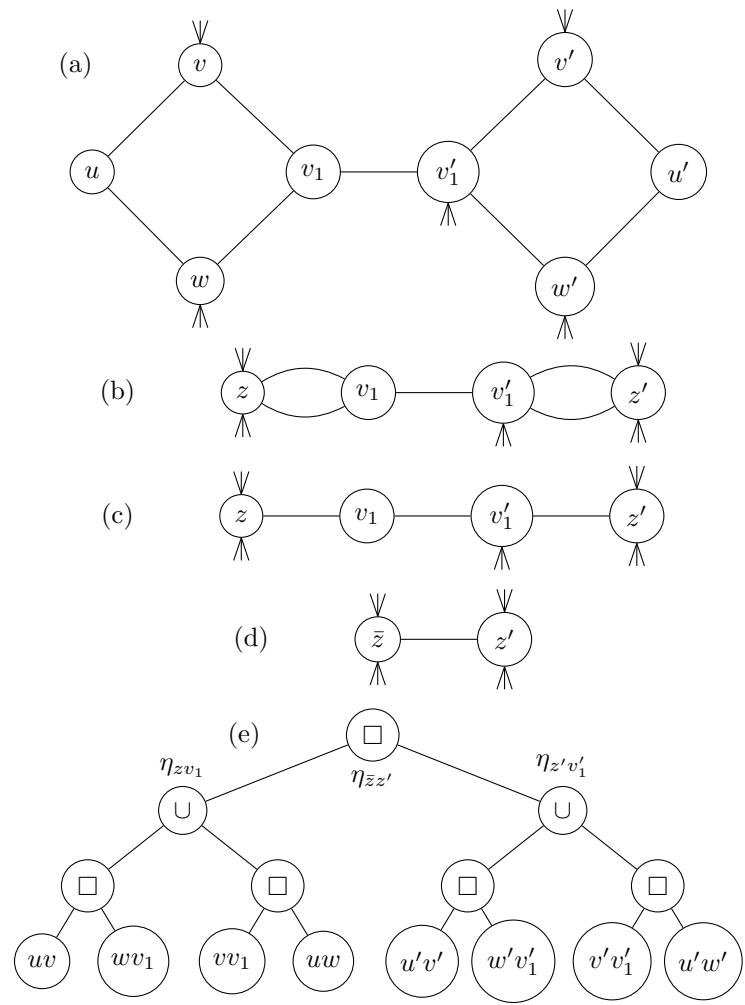

\centering
\include{trim}
\caption{
Figure (a) shows an example of a graph undergoing step-by-step trimming in figures (b) and (c), resulting in the trimmed graph in (d).
A claw attached to a vertex indicates that the vertex has additional neighbors not shown.
Figure (e) depicts the union-product circuit corresponding to the edge $\bar{z}z'$.
}
\label{fig:trimming}
\end{figure}

Handling parallel edges becomes more intricate when they are involved in the trimming of another vertex of degree two.
Figure~\ref{fig:trimming}(a) illustrates this scenario with two vertices with degree 2, $u$ and $u'$.
Their respective contractions lead to the formation of parallel edges in Figure~\ref{fig:trimming}(b).
After removing the redundant parallel edges, the degree of vertex $v_1$ drops to two in Figure~\ref{fig:trimming}(c), prompting another contraction of vertices $z$ and $v_1'$ into a new vertex $\bar{z}$ in Figure~\ref{fig:trimming}(d).
The resulting edge $\bar{z}z'$ in $G'$ represents four distinct perfect matchings in $G$, each containing four edges of $G$.
Since the number of perfect matchings represented by a single edge may grow exponentially, we employ the union-product circuit to represent them compactly.

We say that a graph $G$ is \emph{trimmed} if it satisfies the following conditions.
\begin{itemize}
    \item The graph contains at most one isolated edge $ab$, i.e., $|b^+(G)| \le 1$.
    \item Every component of $G$ except possibly $ab$ is strongly connected; that is, $b^-(G) = \emptyset$.
    \item Every vertex of $G$, except $a$ and $b$, has degree at least three.
\end{itemize}

The trimming algorithm is overviewed later in this section.

\medskip
\textbf{Union-product circuit.}

In the description of the union-product circuit, we distinguish between the input graph $G^\star$, for which all perfect matchings are enumerated, and an intermediate graph $G$. 
We also use $G$ in auxiliary lemmas.

Let $X$ and $Y$ be two sets of matchings such that every matching in $X$ is disjoint from every matching in $Y$.
We define their \emph{product}, denoted $X \Square Y$, as the set of matchings:
$$X \Square Y = \left\{ M_X \dotcup M_Y \;\middle|\; M_X \in X,\ M_Y \in Y \right\}.$$
This operation differs from the Cartesian product, which yields pairs $(M_X, M_Y)$ rather than unions.
The operator $\Square$ is commutative and associative, and satisfies $|X \Square Y| = |X| \cdot |Y|$.
For $k$ sets of pairwise disjoint matchings $X_1, \ldots, X_k$, we define:
$$\Square_{i=1}^{k} X_i = \left\{ \bigcup_{i=1}^k M_i \;\middle|\; M_i \in X_i \text{ for all } i \in \{1,\ldots,k\} \right\}.$$

The \emph{union-product circuit} is a directed acyclic graph where each leaf node corresponds to a single edge of the input graph $G^\star$ and has no outgoing edges.
Every inner node represents either a union operator $\cup$ or a product operator $\Square$ and has two outgoing edges to its children.

Each edge $e$ in the intermediate graph $G$ has a pointer to a node in the union-product circuit, denoted $\eta_e$.
Initially, each edge $e$ of $G^\star$ is assigned a leaf node corresponding to $e$.

The circuit encodes sets of matchings by interpreting each node as an algebraic operation over sets.
Formally, we denote by $\Upsilon(u)$ the set of matchings encoded in a node $u$ recursively from leaves as follows.
\begin{itemize}
    \item If $u$ is a leaf storing edge $e$, then $\Upsilon(u) = \set{\set{e}}$; i.e. a single matching containing only $e$.
    \item If $u$ is a union node with children $l$ and $r$, then $\Upsilon(u) = \Upsilon(l) \cup \Upsilon(r)$.
    \item If $u$ is a product node with children $l$ and $r$, then $\Upsilon(u) = \Upsilon(l) \Square \Upsilon(r)$.
\end{itemize}

To reason inductively about correctness, we define $\Upsilon(e)$ as the set of matchings $\Upsilon(\eta_e)$ encoded by the node $\eta_e$, and define the set of matchings encoded by the entire graph $G$ as:
$$\MM(G) := \bigcup_{M \in \M(G)} \Square_{e \in M} \Upsilon(e).$$

Initially, for $G = G^\star$, each edge $e$ encodes $\Upsilon(e) = \{ \{e\} \}$, and so $\Square_{e \in M} \Upsilon(e) = \{M\}$ for each $M \in \M(G^\star)$, yielding $\MM(G^\star) = \M(G^\star)$.
Throughout the algorithm, matchings are recursively partitioned using the function \Enumerate, which constructs graphs $G_A$ and $G_B$ such that $\M(G) = \M(G_A) \dotcup \M(G_B)$.
Lemma~\ref{lem:AB_encoded} implies that $\MM(G) = \MM(G_A) \dotcup \MM(G_B)$.
Similarly, when trimming transforms $G$ into $G^T$, we maintain $\MM(G) = \MM(G^T)$ as shown in Lemma~\ref{lem:trim_encoded}.
This ensures that each perfect matching of $G^\star$ is visited exactly once by the algorithm.

Trimming applies two fundamental operations to construct the union-product circuit.
For nodes $l$ and $r$, the product $l \Square r$ creates a new product node with children $l$ and $r$.
Similarly, the union $l \cup r$ creates a new union node with children $l$ and $r$.
Trimming uses these operations to ensure that the set of perfect matching $\MM(G)$ encoded in $G$ is unchanged; see Lemma~\ref{lem:trim_encoded} for formal proves.
\begin{itemize}
    \item Removing edges of $b^-(G)$ preserves $\M(G)$, so $\MM(G)$ as well.
    \item If $G$ contains two isolated edges $e$ and $f$, they are replaced by a single edge $g$ with a new node $\eta_g = \eta_e \Square \eta_f$.
Since every perfect matching in $\M(G)$ must include both $e$ and $f$, this preserves $\MM(G)$.
    \item If $G$ contains two parallel edges $e$ and $f$, they are replaced by a single edge $g$ with $\eta_g = \eta_e \cup \eta_f$.
This ensures that every matching containing $g$ corresponds to one containing either $e$ or $f$, preserving the set $\MM(G)$.
    \item When a vertex $u$ with only neighbors $v$ and $w$ is trimmed, and $v_i$ is a neighbor of $v$, the edge $zv_i$ (after contracting $v$ and $w$ into $z$) is assigned a new node $\eta_{zv_i} = \eta_{uw} \Square \eta_{vv_i}$.
\end{itemize}
This construction of union-product circuits is illustrated in Figure~\ref{fig:trimming}.

\begin{figure}[ht]
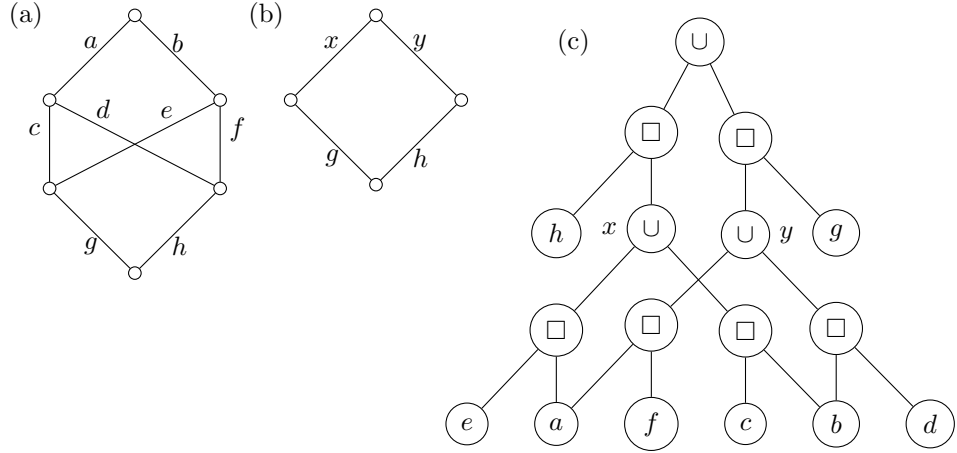

\centering
\include{circuit}
\caption{
An example of a union-product circuit that does not form a tree.
Figure (a) shows a graph where the top vertex is trimmed, resulting in the graph in (b).
Trimming the top vertex again yields the final graph which is a single edge whose node is the top node in (c).
All edges in the circuit are directed downwards.
}
\label{fig:circuit}
\end{figure}

Note that the union-product circuit is not necessarily a tree.
Figure~\ref{fig:circuit}(a) shows a graph that, after successive trimming, results in a single edge whose circuit is shown in Figure~\ref{fig:circuit}(c).
In this case, the leaves storing edges $a$ and $b$ each have two parents.

The algorithm constructs the circuit in a bottom-up manner: every node is created with children that have already been constructed, and once a node is created, it is never modified. This ensures that the set of matchings encoded by each node remains correct and unchanged throughout execution.

\medskip
\textbf{Our algorithm.}

\begin{algorithm}[ht]
\caption{The main recursive function enumerating all perfect matchings.}
\label{alg:main}
\myproc{\Enumerate{A trimmed graph $G$ with its perfect matching}}{
    \eIf{$G$ contains only one edge $e$}{
        \Visit{$\eta_e$}
    }{
        Choose $A, B \subseteq E(G)$ that charge the potential of $G$ using Algorithm \ref{alg:AB}\;
        Let $G_A^T := \Trim(G_A)$ where $G_A = G \setminus A$; see Section \ref{sec:trimming}\;
        \Enumerate{$G_A^T$}\;
        Let $G_B^T := \Trim(G_B)$ where $G_B = G \setminus B$\;
        \Enumerate{$G_B^T$}\;
    }
}
\myproc{\Main{$G^\star$}}{
    Let $G^T := \Trim(G^\star)$\;
    \Enumerate($G^T$)
}
\end{algorithm}

Our main algorithm is presented in Algorithm~\ref{alg:main}.
The input graph $G^\star$ is first processed by the function \Trim, and then the function \Enumerate is invoked on the resulting trimmed graph.

The recursive function \Enumerate receives a trimmed graph $G$ and aims to visit all perfect matchings encoded in $\MM(G)$.
If $G$ consists of a single edge $e$, the recursion terminates, and the function \Visit is called on the node $\eta_e$ associated with $e$, which enumerates all matchings encoded in $\eta_e$.
This base case is correct because $G$ contains exactly one perfect matching consisting of the single edge $e$, implying $\MM(G) = \Upsilon(e)$.

In the general case, the function \Enumerate identifies two disjoint edge sets $A, B \subseteq E(G)$ and constructs subgraphs $G_A = G \setminus A$ and $G_B = G \setminus B$.
Both $G_A$ and $G_B$ are then trimmed, and \Enumerate is recursively invoked on the trimmed versions of $G_A$ and $G_B$.

To ensure that every perfect matching in $G$ is visited exactly once, the edge sets $A$ and $B$ must satisfy $\M(G) = \M(G_A) \dotcup \M(G_B)$.
In this case, we say that the edge sets $A$ and $B$ (or, equivalently, the subgraphs $G_A$ and $G_B$) \emph{split the perfect matchings} of $G$.

Recall that the algorithm operates on a graph $G$ alongside one of its perfect matchings $M$, since constructing the auxiliary directed graph $D(G,M)$ requires such a matching.
The function \Trim is responsible for maintaining the consistency of both $G$ and $M$ during trimming.
Within the function \Enumerate, the matching $M$ belongs to exactly one of $G_A$ or $G_B$.
For the other graph, a new perfect matching is obtained using a directed cycle in the graph $D(G, M)$.

\medskip
\textbf{Potential analysis.}

We use a potential method based on \emph{coins} accounted using \emph{potential function} $\Phi$.
The potential function $\Phi(G)$ of a graph $G$ is defined recursively over a union-product tree structure, starting from the leaves.
\begin{itemize}
  \item The potential of a leaf node $l$ is defined as $\Phi(l) = 1$.
  \item For a union node $u$ with children $l$ and $r$, the potential is given by $\Phi(u) = \Phi(l) + \Phi(r)$.
  \item For a product node $u$ with children $l$ and $r$, the potential is defined as $\Phi(u) = \Phi(l) \cdot \Phi(r)$.
\end{itemize}
Note that $\Phi(u)$ is exactly the number of matchings encoded in the node $u$.
For an edge $e \in E$, we define its potential $\Phi(e)$ as the potential of its associated node $\eta_e$.

The potential of a connected component $H$ of the graph $G^b$ is defined as:
$$\Phi(H) = \Phi_E(H) - |V(H)| + 2, \quad \text{where} \quad \Phi_E(H) = \sum_{e \in E(H)} \Phi(e).$$
The potential of the graph $G$ is then defined as the product of the potentials of all components of $G^b$.
It is important to emphasize that the edges in $b^-(G)$ do not contribute to the potential $\Phi(G)$.

Uno~\cite[Lemma 4]{uno2001fast} proved that every strongly connected graph $H$ has at least $|E(H)| - |V(H)| + 2$ perfect matchings.
This result implies that $\Phi(G)$ is a lower bound on the number of perfect matchings in a bipartite graph $G$. For readers interested in tightness, Observation~\ref{obs:min} presents an example of a strongly connected graph $G$ that has exactly $|E| - |V| + 2$ perfect matchings.

Each visited perfect matching provides two coins, and each coin pays for a constant amount of computation.
The function \Visit called on edge $e$ visits all perfect matchings in $\Upsilon(e)$ for which it obtains $2|\Upsilon(e)|$ coins.
It consumes $|\Upsilon(e)|$ coins for its own work and returns $|\Upsilon(e)|$ coins to its caller \Enumerate.

When \Enumerate is called on a graph $G$, it is responsible for enumerating $\MM(G)$, providing $\Phi(G)$ coins to its own caller, and paying for its own computational cost.
To do this, it obtains $\Phi(G^T_A) + \Phi(G^T_B)$ coins from the recursive calls \Enumerate{$G^T_A$} and \Enumerate{$G^T_B$}.
The sets $A$ and $B$ are chosen such that 
$$\Phi(G_A) + \Phi(G_B) - \Phi(G) \ge c \cdot |E(G)|,$$
for a fixed constant $c = 0.1$.
Lemma \ref{lem:AB} proves that sets $A$ and $B$ can be found in time $\O{|E(G)|}$.
Lemma \ref{lem:trim} shows how to trim the graph $G_A$ into $G^T_A$ in time $\O{|E(G_A)| + \Phi(G^T_A) - \Phi(G_A)}$; and similarly for $B$.
Thus, time complexity of splitting and trimming is
$$\O{|E(G)| + \Phi(G^T_A) - \Phi(G_A) + \Phi(G^T_B) - \Phi(G_B)}$$
which is paid by the coins that remains in the recursion call of \Enumerate:
$$\Phi(G^T_A) + \Phi(G^T_B) - \Phi(G) \ge \Phi(G^T_A) - \Phi(G_A) + \Phi(G^T_B) - \Phi(G_B) + c |E(G)|.$$
Note that Lemma \ref{lem:trim_encoded} that trimming does not decrease the potential.

Accordingly, we say that the edge sets $A$ and $B$ (or the subgraphs $G_A$ and $G_B$) \emph{charge the potential} of $G$ if
\begin{itemize}
    \item $\M(G) = \M(G_A) \dotcup \M(G_B)$ (i.e., perfect matchings are split), and
    \item $\Phi(G_A) + \Phi(G_B) - \Phi(G) \ge c \cdot |E(G)|$.
\end{itemize}

By Lemma~\ref{lem:one_component}, it suffices for the function \Enumerate to focus on computing the sets $A$ and $B$ and performing trimming operations within a single non-trivial component $H$ of $G$.
To ensure that the gain in potential, given by $\Phi(H \setminus A) + \Phi(H \setminus B) - \Phi(H)$, is sufficient to cover the computational cost $\mathcal{O}(|E(H)|)$ of running \Enumerate, we must maintain the vertex partition of $G$ into its components throughout the process.
For clarity and without loss of generality, we assume in the following analysis that $G$ is already trimmed and strongly connected.

\medskip
\textbf{$\M^+$-minimal edge and its properties.}

Let $\M_e^+$ denote the subset of perfect matchings $\M$ of $G$ that contain a given edge $e \in E$, and let $\M_e^-$ denote the subset that do not contain $e$.
Define $G_e^-$ as the graph obtained by removing $e$ from $G$, and $G_e^+$ as the graph obtained by removing all edges incident to $e$.
It follows that $\M_e^- = \M(G_e^-)$ and $\M_e^+ = \M(G_e^+)$, since the edge $e$ is preserved in $G_e^+$ but not in $G_e^-$.

Observe that $f \in b^-(G_e^-)$ if and only if $\M_f^+ \subseteq \M_e^+$.
This equivalence arises from the fact that if $f \in b^-(G_e^-)$, then no perfect matching exists that contains $f$ but not $e$, and thus every perfect matching containing $f$ must also contain $e$.
Similar equivalences hold for the sets $b^-(G_e^+)$, $b^+(G_e^-)$, and $b^+(G_e^+)$.

To construct the graphs $G_A$ and $G_B$ for the recursive step, we first find a $\M^+$-minimal edge.
We say that an edge $\pi \in E$ is $\M^+$-minimal if there exists no edge $e \in E$ such that $\M_e^+$ is a proper subset of $\M_\pi^+$.
To find such an edge, we select $e \in E$ and a perfect matching $M \in \M_e^-$.
If there exists another edge $f$ such that $\M_f^+ \subsetneq \M_e^+$, then Lemma~\ref{lem:M-min-find} describes how to identify a $\M^+$-minimal edge $\pi$ on a directed cycle containing both $e$ and $f$ in $D(G,M)$ in time $\mathcal{O}(|E|)$.
In the remainder of this paper, we assume that $\pi$ is such a $\M^+$-minimal edge.

We now characterize the edges in $b^-(G^-_\pi)$. Lemma~\ref{lem:M-min-bridges} shows that for every edge $e \neq \pi$, we have $e \in b^-(G_\pi^-)$ if and only if $\M_\pi^+ = \M_e^+$.
If $b^-(G_\pi^-) \neq \emptyset$, define $F = b^-(G_\pi^-) \cup \{\pi\}$; that is, $F$ consists of all edges $e$ such that $\M_\pi^+ = \M_e^+$. 
It follows that for every perfect matching $M$ of $G$, every directed cycle in $D(G,M)$ either contains all edges in $F$ or none of them.
Consequently, the components of $G \setminus F$ are strongly connected, and the edges in $F$ interconnect these components into a cycle-like structure.
Recall that the potential of $G^-_\pi$ is the product of the potentials of the components in $G \setminus F$.
This multiplicative structure sufficiently increases the potential, and Lemma~\ref{lem:M-min-multiple} proves that $G_\pi^+$ and $G_\pi^-$ charge the potential of $G$.
In such a case, the recursion proceeds with the graphs $G_\pi^+$ and $G_\pi^-$.
Therefore, for the remainder of the discussion, we assume $b^-(G_\pi^-) = \emptyset$.

Next, consider an edge $uv \in b^+(G^-_\pi)$.
Since $G$ is trimmed and strongly connected, the vertex $u$ has at least two other neighbors, say $x$ and $y$.
Since $\M^+_{ux} \dotcup \M^+_{uy} \subseteq \M^-_{uv} \subseteq \M^+_\pi$ and both $\M^+_{ux}$ and $\M^+_{uy}$ are non-empty, it follows that $\pi$ is not $\M^+$-minimal which is a contradiction; see Lemma~\ref{lem:M-min-bridges} for details.
Thus, we assume $b(G_\pi^-) = \emptyset$, which implies $\Phi(G_\pi^-) = \Phi(G) - \Phi(\pi)$.

However, we cannot proceed with the recursion on $G^+_\pi$ and $G^-_\pi$ if $\Phi(G_\pi^+) - \Phi(\pi) < c \cdot |E(G)|$, which indicates that $b^-(G^+_\pi)$ contains many edges. We now analyze the structure of these edges.

Let $M_\pi$ be a perfect matching of $G$ that contains $\pi$, and let $\sigma$ and $\tau$ be the endpoints of $\pi$, oriented from $\sigma$ to $\tau$ in $D(G, M_\pi)$.
Since $G$ is bipartite, we partition the vertices into $V_\sigma$ and $V_\tau$, with $\sigma \in V_\sigma$ and $\tau \in V_\tau$.

Let $\K$ be the set of all components of $(G_\pi^+ \setminus b^-(G_\pi^+)) \setminus \{\sigma, \tau\}$, i.e., the strongly connected components of $D(G_\pi^+, M_\pi)$ and all trivial components in $b^+(G_\pi^+)$, excluding $\pi$.

We partition the edges of $G$ into three groups:
\begin{itemize}
    \item \emph{Internal edges} $E(\K)$: Edges whose endpoints lie in the same component of $\K$, including trivial components. Formally,
    $$E(\K) = E(G) \setminus (b^-(G^+_\pi) \cup N_E(\tau) \cup N_E(\sigma)).$$
    \item \emph{External edges} $\Ex$: Edges with endpoints in different components of $\K$, excluding $\pi$:
    $$\Ex = (b^-(G^+_\pi) \cup N_E(\tau) \cup N_E(\sigma)) \setminus \{\pi\}.$$
    \item \emph{Edge $\pi$}: Treated as a separate category.
\end{itemize}

Now define a directed graph $\hG$ as follows.
Each component of $\K$ becomes a vertex in $\hG$, and both $\sigma$ and $\tau$ are also vertices in $\hG$.
There is a directed edge from $u$ to $v$ in $\hG$ if and only if $D(G^-_\pi, M_\pi)$ contains a directed edge between the corresponding components.
That is, $\hG$ is obtained by contracting each component of $\K$ into a single vertex, retaining $\sigma$ and $\tau$, and removing parallel edges.

Note that every directed path in $D(G^-_\pi, M_\pi)$ can be shortened to a path in $\hG$ by omitting internal edges of $\K$, and conversely, every path in $\hG$ can be extended into a path in $D(G^-_\pi, M_\pi)$ by reintroducing such edges.
Therefore, $\hG$ is a directed acyclic graph.
For an external edge $uv$, we say that $uv$ is outgoing from $K_u$ and incoming to $K_v$ where $K_u$ and $K_v$ is the component of $\K$ containing $u$ and $v$, respectively, or vertices $\sigma$ or $\tau$.

Define a partial order $<$ on vertices of $\hG$ such that $u < v$ if a directed path from $u$ to $v$ exists in $\hG$.
This relation is reflexive, antisymmetric, and transitive, so $\hG$ can be viewed as a partially ordered set (poset).
The poset has a unique minimal element $\tau$ and a unique maximal element $\sigma$.
Two vertices are incomparable if no directed path exists between them.

A subset $W \subseteq V(\hG)$ is an \emph{antichain} if no two elements of $W$ are comparable.
A subset $I \subseteq V(\hG)$ is an \emph{ideal} if it contains all vertices smaller than each of its elements.
Let $\delta(I)$ denote the set of external edges outgoing from vertices in $I$ and incoming to vertices in the complement of $I$.

Lemma~\ref{lem:ideal_matching} proves that for every perfect matching $M$ of $G$ that avoids $\pi$, the edges in $M \cap \Ex$ form a directed path in $\hG$.
Therefore, $M \cap \delta(I)$ contains exactly one edge.
It follows that if $\delta(I) \cup \{\pi\}$ is partitioned into two disjoint sets $A$ and $B$, then every perfect matching of $G$ is contained in either $G_A$ or $G_B$.
By convention, we always assign $\pi$ to $A$, and we require that $B$ is non-empty to ensure $G_A$ contains at least one perfect matching.

In this paper, we consider only those sets $A$ and $B$ that split $\delta(I) \cup \{\pi\}$ for some ideal $I$.
Algorithm~\ref{alg:AB} finds such sets $A$ and $B$ that also charge the potential.
A frequently used special case splits the edges incident to the vertex $\tau$.

\medskip
\textbf{Components of $G^b_A$ and $G^b_B$.}

To compute the potentials $\Phi(G_A)$ and $\Phi(G_B)$, we must determine which edges of $G$ belong to $b^-(G_A)$ and $b^-(G_B)$, as well as the component structure of $G_A \setminus b^-(G_A)$ and $G_B \setminus b^-(G_B)$, denoted by $G^b_A$ and $G^b_B$, respectively.

An external edge $e \in \Ex \setminus A$ is said to be \emph{reachable} in $G_A$ if there exists a directed path from $\tau$ to $\sigma$ in $D(G, M_\pi) \setminus A$ that includes $e$.
A component $K \in \K$ is \emph{reachable} in $G_A$ if there exists such a path passing through at least one vertex of $K$.
A component $K$ is \emph{avoidable} in $G_A$ if it is reachable in $G_A$, and there also exists a directed path from $\tau$ to $\sigma$ in $D(G, M_\pi) \setminus A$ that avoids all vertices of $K$.
Reachability in $G_B$ is defined analogously.
The negations of these terms are referred to as \emph{unreachable} and \emph{unavoidable}. 
Since edges in $A$ are removed from $G_A$, they are neither reachable nor unreachable in $G_A$.

Let $R_A$ and $R_B$ denote the sets of external edges that are reachable in $G_A$ and $G_B$, respectively, and let $R'_A$ and $R'_B$ denote those that are unreachable.
Similarly, let $\K_A$ and $\K_B$ denote the sets of components of $\K$ that are reachable in $G_A$ and $G_B$, and let $\K'_A$ and $\K'_B$ denote those that are unreachable.

Lemmas~\ref{lem:GA} and~\ref{lem:GB} establish the following properties of $G_A$ and $G_B$.
An external edge $e$ belongs to $b^-(G_A)$ (resp., $b^-(G_B)$) if and only if $e$ is unreachable in $G_A$ (resp., $G_B$).
Furthermore, $b^-(G_B)$ contains only external edges that are unreachable in $G_B$.
As a result, the graph $G^b_B$ consists of all components of $\K$ that are unreachable in $G_B$, along with a single component, denoted by $K^\pi_B$, which contains $\pi$, all external edges and components that are reachable in $G_B$.

\begin{figure}
\centering
\includegraphics[scale=0.7]{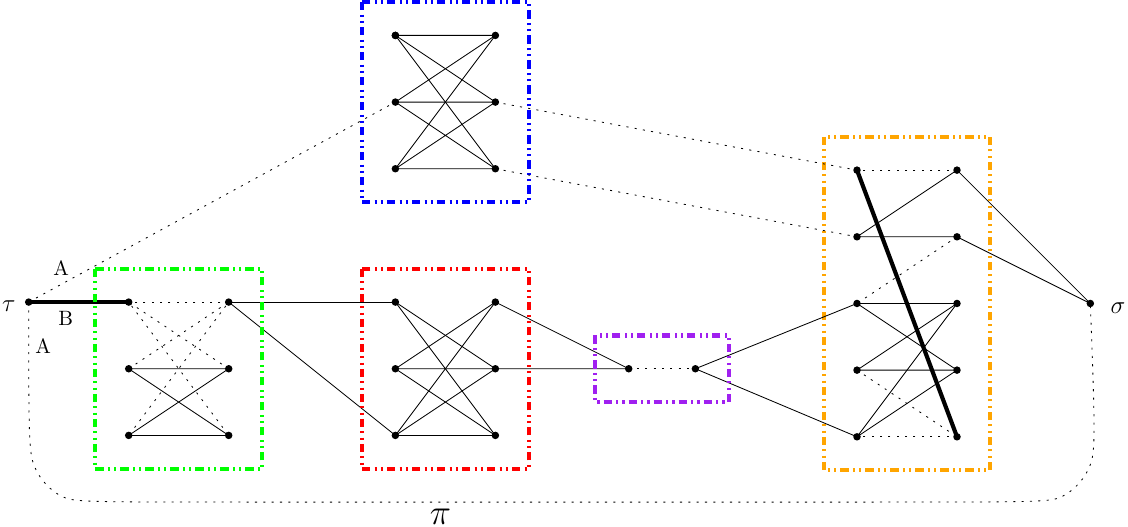}
\caption{Example of a graph $G_A$ with four unavoidable components (green, red, purple, orange), one of which is trivial (purple), and one unreachable component (blue) in $G_A$.
Three edges incident to $\tau$ are split into $A$ and $B$.
Dotted lines represent edges in $b^-(G_A) \cup A$, bold lines represent edges in $b^+(G_A)$, and normal lines represent edges in $G^b_A$.}
\label{fig:unavoidable}
\end{figure}

Internal edges of components that are unreachable or avoidable in $G_A$ do not belong to $b^-(G_A)$.
However, components that are unavoidable in $G_A$ may still contain some edges that do belong to $b^-(G_A)$; see Figure~\ref{fig:unavoidable}.

Let $K$ be a component of $\K$ that is unavoidable in $G_A$.
A vertex of $K$ is called an \emph{input vertex} (resp., \emph{output vertex}) in $G_A$ if it has at least one incoming (resp., outgoing) external edge that is reachable in $G_A$.
Lemma~\ref{lem:unavoidable_GA} shows that all input vertices of $K$ lie in the same component of $K \setminus b^-(G_A)$, and similarly for output vertices, although the input and output components may differ.
In Figure~\ref{fig:unavoidable}, only the red component has input and output vertices lying in the same component of $G^b_A$.

Moreover, if a non-trivial component $K \in \K$ unavoidable in $G_A$ has only one input and one output vertex, then $G^b_A$ contains a non-trivial component entirely contained within $K$, as illustrated by the green component in Figure~\ref{fig:unavoidable}.
On the other hand, if $K$ has multiple input or output vertices, it may contain no non-trivial component in $G^b_A$, or such a component may be trivial, as shown by the orange component.

These components together with non-trivial components unreachable in $G_A$ and $G_B$ are important, as their potentials multiply the potential of all external edges reachable in $G_A$ and $G_B$.
These structural properties are essential in Lemmas~\ref{lem:phi_GA} and Lemma \ref{lem:phi_GB} to derive lower bounds on $\Phi(G_A)$ and $\Phi(G_B)$.

For interested readers, we note that $b^+(G_A)$ contains:
\begin{itemize}
    \item All trivial components of $\K$ that are unreachable in $G_A$,
    \item All external edges that are unavoidable in $G_A$,
    \item Possibly some edges within components of $\K$ that are unavoidable in $G_A$.
\end{itemize}
In contrast, $b^+(G_B)$ contains only the trivial components of $\K$ that are unreachable in $G_B$.

\medskip
\textbf{Choosing sets $A$ and $B$.}
\begin{algorithm}[ht]
\caption{Algorithm constructing sets $A$ and $B$ that charge the potential of $G$}
\label{alg:AB}
Find $\M^+$-minimal edge $\pi = \sigma\tau$ using Lemma~\ref{lem:M-min-find}\;
\If{$\Phi(G^+_\pi) + \Phi(G^-_\pi) - \Phi(G) \ge c|E(G)|$ (Case 1)}{
    \Return sets $A = \{\pi\}$ and $B = N_E(\tau) \setminus \{\pi\}$
}
\If{there exist two non-trivial and incomparable components $K_A, K_B \in \K$ (Case 2)}{
    \Return sets $A,B$ created by Lemma~\ref{lem:antichain}
}
\If{there are at least two edges from $\tau$ incoming to $\kappa_1$ (Case 3)}{
    \Return sets $A,B$ created by Lemma~\ref{lem:K1}\;
    \tcp{Similarly if there are at least two edges outgoing from $\kappa_{\zeta}$ to $\sigma$}
}
\If{$\kappa_1$ has only one outgoing edge (Case 4)}{
    \Return sets $A,B$ created by Lemma~\ref{lem:one_one}\;
    \tcp{Similarly if there is only one edge incoming to $\kappa_{\zeta}$}
}
\If{$\kappa_2$ has only one incoming edge (Case 5)}{
    \Return sets $A,B$ created by Lemma~\ref{lem:K2}\;
    \tcp{Similarly if there is only one edge outgoing from $\kappa_{\zeta-1}$}
}
\Return sets $A,B$ created by Lemma~\ref{lem:AB} (Case 6)
\end{algorithm}

We present Algorithm~\ref{alg:AB}, which finds sets $A, B \subseteq E$ that charge the potential in time $\O{|E(G)|}$.
Recall that $G$ is trimmed and connected.

\smallskip
\textit{Case 1: $G^+_\pi$ and $G^-_\pi$ charge the potential.}
The algorithm begins by finding a $\M^+$-minimal edge $\pi = \sigma\tau$.
If the sets $A = \{\pi\}$ and $B = N_E(\tau) \setminus \set{\pi}$ charge the potential, they are returned immediately.

In the rest of Algorithm \ref{alg:AB}, we assume these sets do not charge the potential, which by Lemma~\ref{lem:M-min-bridges} implies $b(G^-_\pi) = \emptyset$.
Furthermore, since $\Phi(G^+_\pi) - \Phi(\pi) < c|E|$, this indicates that many external edges exist; see Lemma \ref{lem:phi_G_pi} for precise bounds.
Thus, we aim to construct sets $A$ and $B$ such that most external edges are accounted twice in $\Phi(G_A) + \Phi(G_B)$.

To achieve this, we combine two approaches:
\begin{description}
    \item[Unreachable components] (Lemma~\ref{lem:phi}): If both $G_A$ and $G_B$ have non-trivial unreachable components, their potentials multiply contributions from all reachable external edges, ensuring that $A$ and $B$ charge the potential.
    \item[Reachability of external edges] (Lemma~\ref{lem:two_edges}): We find an ideal $I$ of $\hG$ such that every maximal vertex of $I$ has at least two outgoing edges, and every minimal vertex of its complement has at least two incoming edges.
    Then, edges in $\delta(I) \cup \set\pi$ can be split such that all components and external edges are reachable in both $G_A$ and $G_B$.
    If the size of $\delta(I)$ is small, this suffices to charge the potential.
\end{description}

\smallskip
\textit{Case 2: Incomparable Non-Trivial Components.}
From the poset structure of $\hG$, we check whether there exist two non-trivial and incomparable components, denoted by $K_A$ and $K_B$.
If so, let $I$ be the ideal whose maximal elements are $K_A$ and $K_B$, and split the edges in $\delta(I) \cup \{\pi\}$ such that $B$ contains all edges outgoing from $K_B$.
Here, $K_A$ is unreachable in $G_A$ and $K_B$ in $G_B$. Lemma~\ref{lem:antichain} confirms that such sets $A$ and $B$ charge the potential.

Therefore, we assume that every pair of non-trivial components is comparable.
So, $\hG$ contains a maximal chain on vertices $\tau, \kappa_1, \ldots, \kappa_\zeta, \sigma$ that includes all non-trivial components (possibly with some trivial ones).
Since trivial components have at least two incoming and two outgoing edges, the components $\kappa_1$ and $\kappa_\zeta$ are non-trivial and comparable to all vertices of $\hG$.

\textit{Case 3: Two edges from $\tau$ to $\kappa_1$.}
If there are at least two edges from $\tau$ to $\kappa_1$, we construct $B$ to contain only one such edge and assign all other edges incident to $\tau$ to $A$.
Then, all external edges and components of $\K$ are reachable in both $G_A$ and $G_B$, and the total size $|A \cup B| \le |V|/2$.
Lemma~\ref{lem:K1} confirms these graphs charge the potential.

We assume that there is only one edge from $\tau$ to $\kappa_1$.
As $\tau$ must have at least three neighbors, and one is $\sigma$ and one belongs to $\kappa_1$, $\K$ must contain at least two components.
If it contains only two, then $\Phi(G^+_\pi)$ equals the product of their potentials, and Lemma~\ref{lem:zeta_2} guarantees that $G^+_\pi$ and $G^-_\pi$ charge the potential.
Thus, assume $\zeta \ge 3$.

\smallskip
\textit{Case 4: Unique Edge from $\kappa_1$.}
If $\kappa_1$ has only one outgoing edge $e$ (necessarily incoming to $\kappa_2$), we define:
$$B = \{e\}, \quad A = (\delta(\{\tau, \kappa_1\}) \setminus B) \cup \{\pi\}.$$
Here, $\kappa_1$ is unreachable in $G_B$ and unavoidable in $G_A$, and has only one input and one output vertex in $G_A$. By Lemma~\ref{lem:unavoidable_GA}, $G_A$ contains a non-trivial component entirely within $\kappa_1$, and Lemma~\ref{lem:one_one} shows that $G_A$ and $G_B$ charge the potential.

\smallskip
\textit{Case 5: Unique Edge into $\kappa_2$.}
We apply the same construction if $\kappa_2$ has only one incoming edge.
In this case, $\kappa_2$ is unreachable in $G_B$, and $\kappa_1$ is unavoidable in $G_A$ with only one input and one output.
Lemma~\ref{lem:K2} confirms that $G_A$ and $G_B$ again charge the potential.

\smallskip
\textit{Case 6: Final construction.}
By symmetry, we may assume there is only one edge from $\kappa_\zeta$ to $\sigma$, at least two incoming edges to $\kappa_\zeta$, and at least two outgoing edges from $\kappa_{\zeta-1}$.
Lemma~\ref{lem:AB} considers two ideals:
$$I_1 = \{\tau, \kappa_1\}, \quad I_2 = V(\hG) \setminus \{\sigma, \kappa_\zeta\}.$$
Both ideals satisfy the assumptions of Lemma~\ref{lem:two_edges}, so for each, we construct sets $A$ and $B$ such that all external edges and components of $\K$ are reachable in both $G_A$ and $G_B$.

If $\delta(I_1)$ and $\delta(I_2)$ are disjoint, Lemma~\ref{lem:AB} shows that splitting the smaller of these into $A$ and $B$ suffices to charge the potential.
If they intersect, the shared edge is used to split $\delta(I_1)$ in such a way that all components of $\K$ remain reachable in $G_A$ and $G_B$ and are avoidable in $G_A$.
Here, all edges of $G$ except those in $\delta(I_1) \cup \{\pi\}$ contribute to both $\Phi(G_A)$ and $\Phi(G_B)$, which is sufficient to charge the potential.
See Lemma~\ref{lem:AB} for the full analysis.

\medskip
\textbf{The trimming algorithm.}

We describe how to trim a graph $G$ into $G^T$ in time $\mathcal{O}(|E(G)| + \Phi(G^T) - \Phi(G)).$
This trimming operation starts by removing all edges in $b^-(G)$, and merging all edges in $b^+(G)$ into a single isolated edge.
Then, each component is processed independently; hence, we assume that $G$ is strongly connected.

A naive vertex-by-vertex trimming of vertices with degree 2 using Lemma~\ref{lem:trim_encoded} results in $\mathcal{O}(|V|^2)$ time, which is inefficient.
Instead, trimming is carried out in two phases:

\textit{Phase 1.}  
Let $W$ be the set of all vertices with degree 2 in one partite class of $G$.
By Lemma~\ref{lem:trim_forest}, if $N_E(W)$ contains a cycle, then $G$ is a cycle, handled separately by Lemma~\ref{lem:trim_cycle}.
Otherwise, $N_E(W)$ is a forest, and Lemma~\ref{lem:trim_first} trims each tree in $\mathcal{O}(|E|)$ total time.
We then process the vertices with degree 2 in the other partite class in the same way.

\textit{Phase 2.}
Lemma~\ref{lem:trim_potential} shows that after Phase 1, every vertex with degree 2 is incident to an edge with potential at least 2.
This potential is used in Lemma \ref{lem:trim} to prove that the second phase can be processed in time $\mathcal{O}(|E(G)| + \Phi(G^T) - \Phi(G))$.

\medskip
\textbf{Visiting trees of the union-product circuit.}

The function \Visit{$e$}, presented in Algorithm~\ref{alg:visit}, is called by \Enumerate{$G$} when $G$ contains only one edge $e$.
Its goal is to enumerate all perfect matchings in $\Upsilon(\eta_e)$.

As described in the introduction, a perfect matching $M$ in $G^\star$ is represented by a visiting tree: A binary tree with $|M|$ leaves, each corresponding to an edge in $M$, and $|M| - 1$ internal nodes together forming a binary tree.
The visiting tree includes both children of every product node, and exactly one child of each union node.
The other child of a union node, and all nodes unreachable via these visiting pointers from $\eta_e$, are excluded from the visiting tree.

To ensure that all internal nodes in the visiting tree have two children, union nodes are ``skipped'': product nodes directly reference their nearest product descendants.
The leaves of the visiting tree thus correspond exactly to the edges in one perfect matching.

Lemma~\ref{lem:visit_matching} proves a bijection between visiting trees rooted at $\eta_e$ and the matchings in $\Upsilon(\eta_e)$.
The leaves of these trees yield all matchings in $\Upsilon(\eta_e)$.
Furthermore, Lemma~\ref{lem:visit_tree} describes how all visiting trees can be enumerated, and Lemma~\ref{lem:visit_time} proves that this enumeration runs in time $\O{|\Upsilon(\eta_e)|}$.

\medskip
\textbf{Summary.}
\vspace{-3mm}
\begin{proof}[Proof of Theorem \ref{thm:main}]
Algorithm~\ref{alg:main} integrates three key components: recursive splitting (Algorithm~\ref{alg:AB}), graph trimming (Section~\ref{sec:trimming}), and visiting trees (Algorithm~\ref{alg:visit}).
By Lemma~\ref{lem:trim}, trimming preserves the set of encoded perfect matchings. 
The recursive step creates two subgraphs $G_A$ and $G_B$ of a graph $G$ such that the set of perfect matchings encoded in $G$ is split between $G_A$ and $G_B$; see Lemma~\ref{lem:AB_encoded}.
Recursion terminates when a graph consists of a single edge $e$, in which case all matchings encoded in the node $\eta_e$ of the union-product circuit are enumerated via Algorithm~\ref{alg:visit}, as proven by Lemma~\ref{lem:visit_tree}.
Consequently, every perfect matching in the input graph is visited exactly once.

By Lemma~\ref{lem:visit_time}, the total time spent in Algorithm~\ref{alg:visit} is $\O{\M(G^\star)}$.
Furthermore, we conceptually assign $|\M(G^\star)|$ coins to recursive calls of the function \Enumerate which are distributed using the potential $\Phi$.
Each non-terminal call to \Enumerate obtains a trimmed graph $G$, and it performs the following on a selected component $H$ of $G$:
\begin{itemize}
    \item It finds sets $A$ and $B$ of edges in $H$ that charge the potential, using Algorithm~\ref{alg:AB} (based on Lemma~\ref{lem:AB}) in time $\mathcal{O}(|E(H)|)$.
    By Lemma~\ref{lem:one_component}, the potential gain satisfies $\Phi(G_A) + \Phi(G_B) - \Phi(G) \ge c \cdot |E(H)|.$
    \item The graph $H_A = H \setminus A$ is then trimmed into $H_A^T$, in time $\O{|E(H)| + \Phi(H_A^T) - \Phi(H_A)}$, and similarly $H_B$ is trimmed into $H_A^T$; see by Lemma~\ref{lem:trim}.
\end{itemize}

The function \Enumerate receives $\Phi(G_A^T) + \Phi(G_B^T)$ coins from its recursive calls and must pass on $\Phi(G)$ coins to its parent call.
The remaining surplus, $\Phi(G_A^T) + \Phi(G_B^T) - \Phi(G)$ is sufficient to pay for the trimming time $\O{(\Phi(H_A^T) - \Phi(H_A)) + (\Phi(H_B^T) - \Phi(H_B)) + |E(H)|}$, and the cost of computing $A$ and $B$, which is $\mathcal{O}(|E(H)|)$.
Including the initialization of the adjacency matrix and finding the first perfect matching, the overall running time of the algorithm is $\O{|\M| + |V|^2 + |E|^{1+o(1)}}$.

For the space complexity, the adjacency matrix requires $\mathcal{O}(|V(G^\star)|^2)$ space.
All other data structures, including edge lists, components, and node pointers, use $\mathcal{O}(|E(G^\star)|)$ space.
During execution, memory is needed mainly for the union-product circuit.
By Lemma~\ref{lem:trim}, for each removed edge and vertex, $\mathcal{O}(1)$ and $\mathcal{O}(|V(G^\star)|)$ nodes are created, respectively.
As each vertex can be involved in at most one such operation, the total number of created nodes is bounded by $\mathcal{O}(|V(G^\star)|^2)$.
\end{proof}


\section{Preliminaries}

The following observation motivates the definition of the potential $\Phi$.

\begin{figure}[ht]
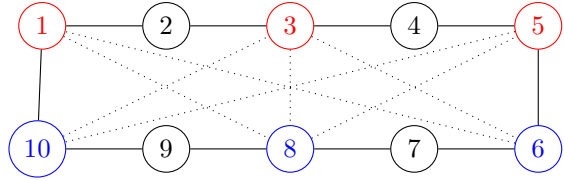

\centering
\include{min}
\caption{
A strongly connected graph with the minimal number of perfect matchings.
Red vertices belong to $A$, and blue vertices belong to $B$.
Solid lines form the cycle $C$, and dotted lines form the set $D$.
}
\label{fig:min}
\end{figure}

\begin{observation} \label{obs:min}
Let $n$ and $m$ be integers such that $n+2$ is divisible by $4$ and $n \le m \le \frac{1}{16}n^2 + \frac{5}{4}n - \frac{7}{4}$.
Then, there exists a strongly connected graph $G = (V, E)$ with $n$ vertices and $m$ edges that contains exactly $|E| - |V| + 2$ perfect matchings.
\end{observation}

\begin{proof}
Let $V = \set{1, \ldots, n}$ be the vertex set, where the odd-numbered vertices form one partite set and the even-numbered vertices form the other in the bipartite graph $G$.
Let $C$ be a cycle on the vertices $1, 2, \ldots, n, 1$.
Define $A$ as the set of all odd vertices in $V$ that are at most $n/2$, and $B$ as the set of all even vertices in $V$ that are at least $n/2$.
Let $D = \set{ ab;\; a \in A,\; b \in B }$; see Figure~\ref{fig:min}.

For any edge subset $F \subseteq D$, the graph $(V, C \cup F)$ contains exactly $|C \cup F| - |V| + 2$ perfect matchings. This is because each perfect matching contains exactly one edge from $D$, and each edge in $(C \cup F) \cap D$ appears in exactly one perfect matching.
The cycle $C$ guarantees that the graph $G$ remains strongly connected for every such subset $F$.
\end{proof}

We say that a graph $G$ \emph{properly encodes} the perfect matchings of an input graph $G^\star$ if the following conditions are satisfied for every node $u$ with children $l$ and $r$:

\begin{enumerate}
    \item Every set of $\Upsilon(u)$ is a matching of $G^\star$.
    \item If $u$ is a union node, then $\Upsilon(l)$ and $\Upsilon(r)$ are disjoint.
    \item If $u$ is a product node, then $\bigcup_{M \in \Upsilon(l)} M$ and $\bigcup_{M \in \Upsilon(r)} M$ are disjoint; i.e., every pair of matchings in $\Upsilon(l)$ and $\Upsilon(r)$ are disjoint.
    \item For every perfect matching $M$ of $G$ and every two edges $e$ and $f$ in $M$, the sets $\bigcup_{M \in \Upsilon(\eta_e)} M$ and $\bigcup_{M \in \Upsilon(\eta_f)} M$ are disjoint.
    \item For any two perfect matchings $M_1$ and $M_2$ of $G$, the sets $\Square_{e \in M_1} \Upsilon(e)$ and $\Square_{e \in M_2} \Upsilon(e)$ are disjoint.
    \item Every set of $\MM(G)$ is a perfect matching of $G^\star$.
\end{enumerate}

Observe that these conditions imply that $\Upsilon(u)$ and $\MM(G)$ are well-defined.
In the initialization of our algorithm, each edge $e$ of $G^\star$ is assigned a leaf storing $e$, satisfying all the conditions above.
We prove that every step of our algorithm maintains these properties.

\begin{lemma} \label{lem:AB_encoded}
If $G$ properly encodes the perfect matchings of $G^\star$ and $\M(G) = \M(G_A) \dotcup \M(G_B)$, then both $G_A$ and $G_B$ properly encode the perfect matchings of $G^\star$, and $\MM(G) = \MM(G_A) \dotcup \MM(G_B)$.
\end{lemma}

\begin{proof}
Since $G_A$ and $G_B$ are subgraphs of $G$, and the union-product circuit remains unchanged, all encoding conditions are preserved.
For each matching $M \in \MM(G)$, there exists a unique perfect matching $M' \in \M(G)$ such that $M \in \Square_{e \in M'} \Upsilon(e)$.
Moreover, $M'$ is a perfect matching either of $G_A$ or $G_B$.
Thus, the perfect matchings of $\MM(G)$ are partitioned into $\MM(G_A)$ and $\MM(G_B)$ according to the partition of $\M(G)$.
\end{proof}

\begin{lemma} \label{lem:trim_encoded}
Let $G$ properly encode the perfect matchings of $G^\star$, and let $G^T$ be obtained by trimming $G$.
Then $G^T$ also properly encodes the perfect matchings of $G^\star$, and $\MM(G) = \MM(G^T)$.
The number of newly created nodes in the union-product circuit is $$\O{|V(G)| (|V(G)| - |V(G^T)| + |E(G)| - |E(G^T)|)},$$
 and trimming a single vertex of degree $2$ with neighbors $v$ and $w$ takes time $\O{\deg(v) + \deg(w)}$.
Moreover, $\Phi(G^T) \ge \Phi(G)$.
\end{lemma}

\begin{proof}
Removing edges from $b^-(G)$ preserves the encoding conditions and set $\MM(G)$ unchanged.

Let $e$ and $f$ be two edges from $b^+(G)$ replaced by a single edge $g$ with a new product node $\eta_g = \eta_e \Square \eta_f$.
Since every perfect matching contains both $e$ and $f$, the fourth condition ensures the third condition holds for $\eta_g$.
Because each matching in $\Upsilon(\eta_g)$ is formed by combining matchings from $\Upsilon(\eta_e)$ and $\Upsilon(\eta_f)$, the remaining conditions also follow, and $\MM(G)$ remains unchanged.

Now, let $u$ be a vertex with only neighbors $v$ and $w$.
Let $G'$ by obtained from $G$ by removing $u$ and edges $uv$, $uw$, and contracting $v$ and $w$ into a new vertex $z$, preserving parallel edges.
For each edge $vy \in E(G)$ with $y \neq u$, $G'$ contains an edge $zy$ with product node $\eta_{zy} = \eta_{vy} \Square \eta_{uw}$, and similarly for edges incident to $w$.
The third condition for each new node follows from the fourth condition applied to perfect matchings of $G$ containing $vy$ (and $uw$).

If $v$ and $w$ have a common neighbor $y \ne u$, $G'$ contains two parallel edges $e$ and $f$ between $z$ and $y$.
These are replaced by a new edge $g$ with a union node $\eta_g = \eta_e \cup \eta_f$.
The second condition for $\eta_g$ follows from the fifth condition applied to two perfect matchings, one containing $e$ and the other $f$.
No perfect matching contains both $e$ and $f$, preserving all conditions and $\MM(G) = \MM(G')$.

Trimming $u$ creates $\deg(v) + \deg(w) - 2$ new nodes in time $\O{\deg(v) + \deg(w)}$.
Merging two parallel or isolated edges adds one new node.
From the definition of the potential it follows that trimming never decreases the potential.
\end{proof}

The following lemma shows that the recursive function \Enumerate can restrict its attention to a single component of the graph.

\begin{lemma} \label{lem:one_component}
Let $G$ be a bipartite graph with component $H$, and let $A$ and $B$ be sets of edges in $H$.
Define $H_A = H \setminus A$ and $H_B = H \setminus B$.
If $\M(H) = \M(H_A) \dotcup \M(H_B)$, then:
$$\M(G) = \M(G_A) \dotcup \M(G_B)
\quad \text{and} \quad
\Phi(G_A) + \Phi(G_B) - \Phi(G) \ge \Phi(H_A) + \Phi(H_B) - \Phi(H).$$
\end{lemma}

\begin{proof}
We first show that $\M(H) = \M(H_A) \dotcup \M(H_B)$ implies $\M(G) = \M(G_A) \dotcup \M(G_B)$.
Let $M$ be a perfect matching of $G$, and let $M_H = M \cap E(H)$, which is a perfect matching of $H$ since $H$ is a component of $G$.
Then, either $M_H \in \M(H_A)$ or $M_H \in \M(H_B)$, meaning that $M$ is a perfect matching of either $G_A$ or $G_B$.

Let $G_1, \ldots, G_k$ be all other components of $G$ (excluding $H$).
By the definition of potential,
$$\Phi(G_A) + \Phi(G_B) - \Phi(G) = (\Phi(H_A) + \Phi(H_B) - \Phi(H)) \cdot \prod_{i=1}^k \Phi(G_i) \ge \Phi(H_A) + \Phi(H_B) - \Phi(H).$$
\end{proof}

The next lemma gives a relationship between the product and sum of potentials of graph components.

\begin{lemma} \label{lem:potential_edge}
For every graph $G$, it holds that $\Phi(G) \ge \Phi_E(G) - |V(G)| + \scc(G) + 1,$
where $\scc(G)$ denotes the number of components in $G \setminus b^-(G)$.
\end{lemma}

\begin{proof}
Let $G_1, \ldots, G_k$ be the components of $G$.
Since $\Phi(G) = \prod_{i=1}^k \Phi(G_i)$ and
$$\Phi_E(G) - |V(G)| + k = \sum_{i=1}^k (\Phi(G_i) - 1),$$
we need to prove:
$$\prod_{i=1}^k \Phi(G_i) \ge 1 + \sum_{i=1}^k (\Phi(G_i) - 1).$$
As $\Phi(G_i) \ge 1$, the inequality is trivial for $k = 1$.
For $k = 2$, we have:
$$0 \le (\Phi(G_1)-1)(\Phi(G_2)-1) = \Phi(G_1)\Phi(G_2) - \Phi(G_1) - \Phi(G_2) + 1.$$
Hence,
$$\Phi(G_1)\Phi(G_2) \ge \Phi(G_1) + \Phi(G_2) - 1.$$
For $k \ge 3$, proceed by induction:
$$\prod_{i=1}^k \Phi(G_i) \ge \Phi(G_1) \cdot \left(1 + \sum_{i=2}^k (\Phi(G_i) - 1)\right) \ge 1 + \sum_{i=1}^k (\Phi(G_i) - 1).$$
\end{proof}


\section{\texorpdfstring{$\M^+$-minimal edge}{M+ minimal edge}}

This section identifies an $\M^+$-minimal edge $\pi$ and presents its fundamental properties.
Recall that find such an edge $\pi$ is the first step of Algorithm \ref{alg:AB}, so we can assume that $G$ is connected and trimmed.

\begin{lemma}\label{lem:M-min-bridges}
Let $\pi$ be an $\M^+$-minimal edge of $G$. Then, for every edge $e \neq \pi$, it holds that $e \in b^-(G_\pi^-)$ if and only if $\M_\pi^+ = \M_e^+$.
Furthermore, $b^+(G_\pi^-)$ contains only isolated edges of $G$.
\end{lemma}

\begin{proof}
Recall that $e \in b^-(G_\pi^-)$ if and only if $\M_e^+ \subseteq \M_\pi^+$.
By the $\M^+$-minimality of $\pi$, the inclusion $\M_e^+ \subseteq \M_\pi^+$ implies that $\M_e^+ = \M_\pi^+$. This proves the first claim.

Clearly, every isolated edge of $G$ belongs to every perfect matching and hence to $b^+(G_\pi^-)$.
For contradiction, suppose $b^+(G_\pi^-)$ contains an edge $e$ that is not isolated in $G$.
Since $G$ is trimmed, one endpoint of $e$ is incident to two other edges, say $f$ and $g$.
As before, $e \in b^+(G_\pi^-)$ if and only if $\M_e^- \subseteq \M_\pi^+$.
Hence, $\M_f^+ \dotcup \M_g^+ \subseteq \M_e^- \subseteq \M_\pi^+.$
Since both $\M_f^+$ and $\M_g^+$ are non-empty, we get $\M_f^+ \subsetneq \M_\pi^+$, which contradicts the minimality of $\pi$.
\end{proof}

\begin{lemma} \label{lem:M-min-find}
An $\M^+$-minimal edge of $G$ can be found in time $\O{|E(G)|}$.
\end{lemma}
\begin{proof}
To find an $\M^+$-minimal edge, we must be able to determine whether a given edge $e$ is $\M^+$-minimal in time $\O{|E|}$.
For a given edge $e$, compute $b^-(G_e^-)$ and $b^+(G_e^+)$ in time $\O{|E|}$ using Lemma~\ref{lem:basic_structure}.
Note that $h \in b^-(G_e^-) \setminus b^+(G_e^+)$ if and only if $\M^+_h \subsetneq \M^+_e$.
Therefore, $e$ is $\M^+$-minimal if and only if $b^-(G_e^-) \subseteq b^+(G_e^+)$, which can be checked in time $\O{|E|}$.

The algorithm starts with an arbitrary edge $e$. If $b^-(G_e^-) \subseteq b^+(G_e^+)$ holds, then $e$ is returned as $\M^+$-minimal.
So, we assume that there exists an edge $h \in b^-(G_e^-) \setminus b^+(G_e^+)$.

Using Lemma~\ref{lem:basic_structure}, decompose $G_e^-$ into the set $b^-(G_e^-)$ and the components of $G_e^- \setminus b^-(G_e^-)$.
Let $M \in \M$ be a perfect matching avoiding $e$ (and hence $h$ as well).
Let $C_h$ be a directed cycle in $D(G,M)$ containing $h$, and hence also $e$.

Traverse $C_h$ from $e$ to find the first edge $\pi \in b^-(G_e^-)$ incomming to a component $K$ of $G_e^- \setminus b^-(G_e^-)$ that has another incoming edge $g \in b^-(G_e^-)$.
We prove that such a component $K$ exists and that $\pi$ is $\M^+$-minimal.

Since $h \notin b^+(G_e^+)$, there exists a perfect matching $M'$ containing $e$ and avoiding $h$, implying a cycle $C'$ in $M \triangle M'$ containing $e$ and avoiding $h$.
Let $z$ be the first vertex on $C_h$ (traversed from $h$ to $e$) that also lies on $C'$, and let $K_z$ be the component of $G_e^- \setminus b^-(G_e^-)$ containing $z$.
The cycles $C_h$ and $C'$ enter $K_z$ via different edges, so $K$ and $\pi$ exists.

Assume for contradiction that $\pi$ is not $\M^+$-minimal; i.e., there exists $f$ such that $\M_f^+ \subsetneq \M_\pi^+$.
Let $C_f$ be a directed cycle in $D(G,M)$ containing $f$. Since $\M_f^+ \subsetneq \M_\pi^+ \subseteq \M_e^+$, the cycle $C_f$ also contains $\pi$ and $e$.

\textbf{Case 1: $\pi$ is reached before $f$ on $C_f$.}
Let $C_g$ be a directed cycle in $D(G,M)$ containing $g$ and hence $e$. Since $\pi$ and $g$ both enter the same component $K$, construct a directed cycle $\bar{C}$ in $D(G,M)$ as follows:
\begin{itemize}
    \item from $e$ to $g$ along $C_g$,
    \item from the edge exiting $K$ to $e$ along $C_f$, and
    \item a path inside $K$ connecting these two parts.
\end{itemize}
Then $\bar{C}$ is a directed cycle $D(G,M)$ in containing $f$ and avoiding $\pi$, implying that $M \triangle \bar{C} \in \M^+_f \setminus \M^+_\pi$, a which is a contradiction to the assumption that $\M^+_f \subseteq \M^+_\pi$.

\textbf{Case 2: $f$ is reached before $\pi$ on $C_f$.}
If $C_h$ avoids $f$, then $C_h$ and $C_f$ both contain $\pi$, but $C_f$ reaches $f$ before $\pi$, contradicting the choice of $K$ as the first such component with two incoming edges on $C_h$.
Therefore, $C_h$ must contain $f$.

Since $\M_f^+ \subsetneq \M_\pi^+$, there exists a cycle $C^\star$ avoiding $f$ and containing $\pi$, and hence $e$.
So, cycles $C_h$ and $C^\star$ contradicts the assumption that $K$ is the first component on $C_h$ with two incoming edges.

Hence, $\pi$ is $\M^+$-minimal, and it can be found in time $\O{|E|}$.
\end{proof}

The essential property of an $\M^+$-minimal edge $\pi$ is that the graph $G_\pi^+$ remains strongly connected, except for one special case handled by the following lemma.

\begin{lemma}\label{lem:M-min-multiple}
Let $\pi$ be an $\M^+$-minimal edge of $G$.
If $b(G_\pi^-) \ne \emptyset$, then $G_\pi^+$ and $G_\pi^-$ charge the potential of $G$.
\end{lemma}

\begin{proof}
Let $F$ be the set of edges $e \in E$ such that $\M_e^+ = \M_\pi^+$, including $\pi$.
The minimality of $\pi$ implies $F = b^-(G_\pi^-) \cup \{\pi\}$.
Moreover, every directed cycle in $D(G,M)$ either includes all edges in $F$ or none.

Let $K_1, \ldots, K_k$ be the components of $G \setminus F$.
These components are interconnected by the edges of $F$ into a cycle structure: contracting each $V(K_i)$ into a vertex gives a cycle of edges in $F$.
Thus, each component $K_i$ has exactly one incoming and one outgoing edge, making all $K_i$ non-trivial with at least 6 vertices and potential at least 4.

Using induction and Lemma~\ref{lem:potential_edge}, we have:
$$\Phi(G_\pi^-) = \prod_{i=1}^k \Phi(K_i) \ge 2 \sum_{i=1}^k \Phi(K_i) \ge 2(\Phi_E(G \setminus F) - |V(G)| + 2k).$$

Since each edge in $F$ forms a trivial component in $G_\pi^+$, Lemma~\ref{lem:potential_edge} gives:
$$\Phi(G_\pi^+) \ge \prod_{e \in F} \Phi(e) \ge \Phi_E(F) - k + 1.$$

Combining:
\begin{align*}
\Phi(G_\pi^+) + \Phi(G_\pi^-) - \Phi(G)
&\ge \Phi_E(G \setminus F) - |V(G)| + 3k - 1 \\
&= |E(G)| - |F| - |V(G)| + 3k - 1 \\
&\ge \frac{1}{3}|E(G)| + 2k - 1 \ge c|E(G)|.
\end{align*}
\end{proof}

For the remainder of the paper, assume $b(G_\pi^-) = \emptyset$.
The following inequalities are useful when $G_\pi^+$ and $G_\pi^-$ do not charge the potential of $G$.

\begin{lemma} \label{lem:phi_G_pi}
If $G_\pi^+$ and $G_\pi^-$ do not charge the potential of $G$, then:
\begin{equation} \label{eq:phi_G_pi_basic}
c|E(G)| > |E(\K)| - |V(G)| + |\K| + 2.
\end{equation}

Furthermore, if $G_\pi^+$ contains at least two non-trivial components $K', K''$, then:
\begin{equation} \label{eq:phi_G_pi_twice}
c|E(G)| > 2 (|E(\K)| - |V(G)| + |\K|) + 3.
\end{equation}

Additionally, if both $K'$ and $K''$ have at most one incoming or at most one outgoing edge, then:
\begin{equation} \label{eq:phi_G_pi_prod}
c|E(G)| > (|E(K')| - |V(K')| + 2)(|E(K'')| - |V(K'')| + 2) - 1 \ge \frac{1}{4} (|V(K')| + 2)(|V(K'')| + 2) - 1.
\end{equation}
\end{lemma}

\begin{proof}
Since $b(G_\pi^-) = \emptyset$, we have $\Phi(G_\pi^-) = \Phi(G) - \Phi(\pi)$.
As $\pi$ forms a trivial component in $G_\pi^+$, Lemma~\ref{lem:potential_edge} implies:
$$\Phi(G_\pi^+) + \Phi(G_\pi^-) - \Phi(G) = \Phi(\K)\Phi(\pi) - \Phi(\pi) \ge \Phi(\K) - 1 \ge |E(\K)| - |V(\K)| + |\K|.$$
This proves \eqref{eq:phi_G_pi_basic}, using $V(G) = V(\K) \cup \{\sigma, \tau\}$.

For \eqref{eq:phi_G_pi_twice}, both $|E(K')| - |V(K')|$ and $|E(\K \setminus K')| - |V(\K \setminus K')| + |\K| - 2$ are non-negative. Hence, by Lemma~\ref{lem:potential_edge}:
\begin{align*}
\Phi(\K) &= \Phi(K')\Phi(\K \setminus K') \\
&\ge (|E(K')| - |V(K')| + 2)(|E(\K \setminus K')| - |V(\K \setminus K')| + |\K|) \\
&\ge 2(|E(\K)| - |V(\K)| + |\K|).
\end{align*}

For \eqref{eq:phi_G_pi_prod}, since each vertex has degree at least 3, we get $|E(K')| \ge \tfrac{3}{2}|V(K')| - 1$, so:
\begin{align*}
\Phi(K') &\ge |E(K')| - |V(K')| + 2 \ge \tfrac{1}{2}|V(K')| + 1, \\
\Phi(K'') &\ge \tfrac{1}{2}|V(K'')| + 1, \\
\Phi(\K) &\ge \Phi(K') \Phi(K'') \ge \tfrac{1}{4} (|V(K')| + 2)(|V(K'')| + 2).
\end{align*}
\end{proof}


\section{\texorpdfstring{Components and potentials of $G_A$ and $G_B$}{Components and potentials of GA and GB}}
\label{sec:bridges}

In this section, we estimate $\Phi(G_A)$ and $\Phi(G_B)$.
To this end, we analyze which edges belong to $b(G_A)$ and $b(G_B)$, and examine the structure of the components in $G^b_A$ and $G^b_B$.

Recall the current setup.
The graph $G$ is trimmed and connected.
The edge $\pi = \sigma\tau$ is $\M^+$-minimal.
From Lemmas \ref{lem:M-min-bridges} and \ref{lem:M-min-multiple}, we know that $b(G_\pi^-) = \emptyset$.
The graph $G$ is decomposed into components $\K$, the edge $\pi$, and the set of external edges $\Ex = (b^-(G^+_\pi) \cup N(\tau) \cup N(\sigma)) \setminus \set\pi = E(G) \setminus (E(\K) \cup \set\pi)$.
The perfect matching $M_\pi$ in $G$ includes $\pi$, implying that $M_\pi$ contains no external edges.

The sets $A$ and $B$ partition the edges $\delta(I) \cup \set\pi$ for some ideal $I$ such that $A$ contains $\pi$.
If $A$ contains only $\pi$, then $G_A = G^+_\pi$ and $G_B = G^-_\pi$.
Since Algorithm~\ref{alg:AB} begins by checking whether $G^+_\pi$ and $G_B$ charge potential, we assume that $A$ includes $\pi$ and at least one edge from $\delta(I)$.
If $B$ is empty, then $G_B = G$ and $G_A$ has no perfect matching, so they do not charge the potential.
Therefore, we also assume that $B$ contains at least one edge from $\delta(I)$.
These assumptions imply that $G_A$ has a perfect matching $M_A$.

\begin{lemma} \label{lem:directed_cycle}
Perfect matchings of $G$ are split into perfect matchings of $G^+_\pi$ that avoid all external edges, and perfect matchings of $G^-_\pi$ that include at least one external edge.
Every directed cycle $C$ in $D(G,M_\pi)$ either contains $\pi$, or is entirely contained within a single component of $\K$.
\end{lemma}
\begin{proof}
Clearly, every perfect matching of $G$ lies either in $G^+_\pi$ or $G^-_\pi$.
Perfect matchings avoiding $\pi$ must include an edge incident to $\pi$, which is external.
Those containing $\pi$ cannot include any other edge incident to $\tau$ or $\sigma$, nor any edge of $b^-(G^+_\pi)$ by definition.

If $\pi \in C$, then $C$ includes two edges incident to $\pi$, which are external. Thus, $C$ cannot be contained in a single component of $\K$.
Conversely, if $C$ is not contained within a component of $\K$, then it contains an external edge, which implies that the matching $M_\pi \triangle C$ contains an external edge and avoids $\pi$, so $\pi \in C$.
\end{proof}

We now justify the significance of the edges $\delta(I)$, outgoing from an ideal $I$ of $\hG$, in splitting the perfect matchings in $G$.

\begin{lemma} \label{lem:ideal_matching}
Let $M$ be a perfect matching of $G$ avoiding $\pi$, and let $I$ be an ideal of $\hG$.
The set of external edges in $M$ forms a directed path in $\hG$ from $\tau$ to $\sigma$, so the ideal $I$ has exactly one outgoing edge in $M$.
Therefore, if $A$ and $B$ split $\delta(I) \cup \set\pi$, then they split the perfect matchings of $G$.
\end{lemma}
\begin{proof}
The symmetric difference $M \triangle M_\pi$ decomposes into disjoint directed cycles in $D(G, M_\pi)$, one of which contains $\pi$, denoted by $C$.
By Lemma \ref{lem:directed_cycle}, all other cycles lie entirely within components of $\K$ and therefore include no external edges.
By contracting all vertices within each component of $\K$ along cycle $C$ and removing $\pi$, we obtain a directed path in $\hG$ from $\tau$ to $\sigma$.
Let $u$ be the last vertex of $C$ in $I$, and let $e$ be the edge leaving $u$.
Then $e$ is the unique edge of $C$ in $\delta(I)$.

Since $A \dotcup B = \delta(I) \cup \set\pi$, this edge $e$ lies in either $A$ or $B$, so $M$ belongs to either $G_A$ or $G_B$.
Perfect matchings containing $\pi$ avoid all external edges, so they belong only to $G_B$, not to $G_A$.
As $G_A$ and $G_B$ are subgraphs of $G$, they cannot contain any perfect matching not already present in $G$.
Thus, $A$ and $B$ indeed split the perfect matchings of $G$.
\end{proof}

Next, we describe the edges belonging to $b^-(G_A)$ and the components of $G^b_A$.
Note that components of $\K$ that are unavoidable in $G_A$ are analyzed in Lemma~\ref{lem:unavoidable_GA}.

\begin{lemma}\label{lem:GA}
The graph $G_A$ has the following properties:
\begin{enumerate}
    \item An external edge $e \notin A$ belongs to $b^-(G_A)$ if and only if $e$ is unreachable in $G_A$.
    \item A trivial component $K \in \K$ is a trivial component in $G^b_A$ if and only if $K$ is unreachable in $G_A$.
    \item If a component $K \in \K$ is unreachable or avoidable in $G_A$, then $b^-(G_A)$ contains no edge of $K$.
    \item Every component of $\K$ that is unreachable in $G_A$ appears as a component in $G^b_A$.
\end{enumerate}
\end{lemma}
\begin{proof}
First, an external edge $e$ is reachable in $G_A$ if and only if there exists a directed path from $\tau$ to $\sigma$ in $D(G, M_\pi) \setminus A$ that includes $e$.
This implies $e \notin b^-(G_A)$.
Conversely, if $e \in \Ex \setminus (b^-(G_A) \cup A)$, then some perfect matching $M$ in $G_A$ includes $e$.
Thus, $M \triangle M_\pi$ consists of directed cycles within components of $\K$ and one directed cycle containing $e$ and $\pi$, by Lemma~\ref{lem:directed_cycle}, so $e$ is reachable in $G_A$.

Second, consider a trivial component $K$ consisting of a single edge $e$.
If $K$ is unreachable in $G_A$, then all edges incident to $e$ are unreachable and hence in $b^-(G_A)$, implying that $e \in b^+(G_A)$.
Conversely, if $K$ is reachable in $G_A$, then some incident edge is reachable, so $e \notin b^-(G_A)$ and thus not in $b^+(G_A)$.

Third, let $K$ be a non-trivial component of $\K$ that is unreachable in $G_A$, and let $e \in E(K)$.
Since $e \notin b(G^+_\pi)$, there exists a directed cycle $C_e$ in $D(G, M_\pi)$ containing $e$ and avoiding $\pi$.
By Lemma~\ref{lem:directed_cycle}, $C_e$ lies entirely within $K$, and therefore avoids all edges in $A$.
Furthermore, $M_A \triangle M_\pi$ contains an alternating cycle $C_\pi$ including $\pi$ and avoiding $A \setminus \{\pi\}$.
Since, $K$ is unreachable in $G_A$, the cycle $C_\pi$ avoids whole component $K$, so $C_\pi$ and $C_e$ are vertex-disjoint.
Therefore, $C_e$ is a directed cycle in $D(G_A, M_\pi \triangle C_\pi)$, implying $e \notin b(G_A)$.

If $K$ is avoidable in $G_A$, then $D(G, M_\pi) \setminus (A \setminus \set{\pi})$ contains a directed cycle $C_\pi$ passing $\pi$ that is vertex-disjoint with $C_e$, again implying $e \notin b(G_A)$.

Fourth, if $K \in \K$ is unreachable in $G_A$, then we already know that no edge of $K$ lies in $b^-(G_A)$ and all incoming and outgoing edges incident to $K$ are in $b^-(G_A)$.
Hence, $K$ appears as a component in $G^b_A$.
\end{proof}

Using the last lemma, we describe the structure of $G_B$ and estimate its potential.

\begin{lemma} \label{lem:GB}
The graph $G_B$ has the following properties:
\begin{enumerate}
    \item No internal edge belongs to $b^-(G_B)$.
    \item An external edge $e \notin B$ belongs to $b^-(G_B)$ if and only if $e$ is unreachable in $G_B$.
    \item The components of $G^b_B$ are all components of $\K$ that are unreachable in $G_B$ and one additional component that contains $\pi$, all external edges, and all components of $\K$ reachable in $G_B$, denoted by $K^\pi_B$.
\end{enumerate}
\end{lemma}
\begin{proof}
Let $B' = B \cup \{\pi\}$ and let $G_{B'} = G \setminus B'$.
Since $B'$ contains $\pi$, Lemma~\ref{lem:GA} applies to $G_{B'}$.
Using the equality $\M(G_B) = \M(G^+_\pi) \cup \M(G_{B'})$, it follows that
$$
b^-(G_B) = b^-(G_{B'}) \cap b^-(G^+_\pi), \quad
b^+(G_B) = b^+(G_{B'}) \cap b^+(G^+_\pi).
$$
Thus, the first statement holds: internal edges do not belong to $b^-(G_B)$, since they do not appear in $b^-(G^+_\pi)$.

Furthermore, since $G_B$ and $G_{B'}$ share the same set of unreachable external edges and components, the second statement follows.

Consequently, the components of $\K$ that are unreachable in $G_B$ appear as components of $G^b_B$, as asserted in the third statements.
All external edges and all components of $\K$ reachable in $G_B$ together with the edge $\pi$ are interconnected in a single strongly connected component in $D(G_B, M_\pi)$, proving the last statement.
\end{proof}

An interested reader may observe that an edge $e$ of $G_B$ belongs to $b^+(G_B)$ if and only if $e$ is a trivial component of $\K$ that is unreachable in $G_B$.

\begin{lemma} \label{lem:phi_GB}
If every component of $\K$ is reachable in $G_B$, then
$$\Phi(G_B) = \Phi(G) - \Phi_E(B).$$
If there exists a non-trivial component of $\K$ that is unreachable in $G_B$, then
$$\Phi(G_B) \ge 2 \left( \Phi_E(G) - \Phi_E(R'_B) - |V(G)| + |\K'_B| + 1 \right),$$
where $R'_B$ is the set of external edges unreachable in $G_B$ and $\K'_B$ is the set of components of $\K$ unreachable in $G_B$.
\end{lemma}
\begin{proof}
If every component of $\K$ is reachable in $G_B$, then $b(G_B) = \emptyset$ and $G^b_B$ is connected, by Lemma~\ref{lem:GB}.
Therefore, the only contribution to the difference $\Phi(G) - \Phi(G_B)$ comes from the removed edges $B$.

Now assume that there exists a non-trivial component in $\K$ that is unreachable in $G_B$.
Then, by Lemmas~\ref{lem:GB} and~\ref{lem:potential_edge}, we obtain:
\begin{align*}
\Phi(G_B) &= \Phi(K^\pi_B) \cdot \Phi(\K'_B) \\
&\ge \left( \Phi_E(K^\pi_B) - |V(K^\pi_B)| + 2 \right) \cdot \Phi(\K'_B) \\
&\ge 2 \left( \Phi_E(K^\pi_B) - |V(K^\pi_B)| \right) + 2 \left( \Phi_E(\K'_B) - |V(\K'_B)| + |\K'_B| + 1 \right) \\
&= 2 \left( \Phi_E(G) - \Phi_E(R'_B) - |V(G)| + |\K'_B| + 1 \right).
\end{align*}
\end{proof}

Now, we turn back to $G_A$ to study its unavoidable components.
Recall, that vertices of a component $K \in \K$ having at least one incoming (outgoing) external edge reachable in $G_A$ are called input (output) vertices of $K$ in $G_A$.

\begin{lemma} \label{lem:unavoidable_GA}
Let $K$ be a non-trivial component of $\K$ that is unavoidable in $G_A$, and let $K^b = K \setminus b^-(G_A)$.
Then:
\begin{enumerate}
    \item All input vertices of $K$ in $G_A$ lie in the same component of $K^b$, and likewise for the output vertices.
    \item If $K$ has exactly one input and one output vertex, then $G^b_A$ contains a non-trivial component $\bar K$ such that $\bar K$ forms a component of $K^b$.
\end{enumerate}
\end{lemma}
Recall that components containing input and output vertices may be different; see Figure \ref{fig:unavoidable}.
\begin{proof}
Let $x_1y_1$ and $x_2y_2$ be two external edges that are reachable in $G_A$ and incoming to two distinct input vertices $y_1$ and $y_2$ of $K$.
There exist directed paths $P_1$ and $P_2$ in $D(G, M_\pi) \setminus A$ from $\tau$ to $\sigma$ such that $x_1y_1 \in P_1$ and $x_2y_2 \in P_2$.
Let $z$ be the last vertex of $P_1$ lying in $K$.
Since $K$ is strongly connected in $D(G, M_\pi)$, there exists a directed path $P'$ from $y_2$ to $z$ within $K$.
Define a directed cycle $C$ as follows: follow $P_2$ from $\tau$ to $y_2$, then continue along $P'$ to $z$, then along $P_1$ to $\sigma$, and return to $\tau$ via the edge $\pi$.

Both $M_\pi \triangle C$ and $M_\pi \triangle (P_1 \cup \{\pi\})$ are perfect matchings of $G_A$, and hence no edge in their symmetric difference $C \triangle P_1$ belongs to $b(G_A)$.
Note that $x_1y_1$ and $x_2y_2$ are the only external edges in $C \triangle P_1$ that are incident to $K$.
Therefore, the path from $y_2$ to $y_1$ in $C \triangle P_1$ lies entirely within $K$ and consists only of edges not in $b(G_A)$.
This implies that $y_1$ and $y_2$ are in the same component of $K^b$.
A similar argument applies to output vertices.

For the second part, suppose that $K$ has exactly one input vertex $y$ and one output vertex $v$.
Since every perfect matching of $G_A$ must contain exactly one external edge incoming to $y$, all edges of $K$ incident to $y$ belong to $b^-(G_A)$; the same holds for vertex $v$.
Thus, $y$ and $v$ are isolated in $K^b$, and every other vertex of $K$ lies in a component of $K^b$ and these components of $K^b$ are also components of $G^b_A$.

It remains to show that one such component is non-trivial.
We construct a directed cycle in $D(G, M_A)$ that lies entirely within $K$, implying the existence of a non-trivial component in $K^b$.

Without loss of generality (by symmetry in $\hG$), assume $K$ does not belong to the ideal $I$ that defines the partition $(A, B)$.
Since $K$ is non-trivial, there exists a vertex $u_1 \in V_\sigma \cap V(K)$ with $u_1 \neq y$.
Because $u_1$ is not the input vertex, it is matched in $M_A$ to some vertex $u_2 \in V_\tau \cap V(K)$ which is not the output vertex.

Since $K$ is reachable in $G_A$, the vertex $u_2$ is not incident to any external edge, so $u_2$ must have at least one neighbor $u_3 \in V(K)$ distinct from $u_1$ and $y$.
Continue this process: the matching edge $u_3u_4$ lies in $M_A$, and $u_4$ is again not an output vertex.
We repeat this construction until we encounter a vertex $u_k$ that is adjacent to some earlier vertex $u_i$ with $1 \le i \le k-3$, thus closing a cycle.
This yields a directed cycle in $D(G, M_A)$ entirely within $K$, proving that $K^b$ contains a non-trivial component.
\end{proof}

\begin{lemma} \label{lem:phi_GA}
Let $t$ be the number of components of $\K$ that are unavoidable in $G_A$.
Then:
$$\Phi(G_A) \ge \Phi_E(R_A) - t \ge \Phi_E(R'_B) - |\K'_B|.$$

Furthermore, if:
\begin{itemize}
    \item there exists a non-trivial component of $\K$ that is unreachable in $G_A$, or
    \item there exists a non-trivial unavoidable component of $\K$ in $G_A$ with exactly one input and one output vertex,
\end{itemize}
then:
$$\Phi(G_A) \ge 2(\Phi_E(R_A) - t) \ge 2(\Phi_E(R'_B) - |\K'_B|).$$
\end{lemma}
\begin{proof}
Let $G'$ be the subgraph of $G^b_A$ that contains all components of $G^b_A$ that have at least one external edge reachable in $G_A$.
Equivalently, we remove from $G^b_A$ all components of $\K$ unreachable in $G_A$ and all components inside components unavoidable in $G_A$.
By this construction, $\Phi(G_A) \ge \Phi(G')$, so it suffices to lower-bound $\Phi(G')$.
For a component $K \in \K$, let $K'$ be the graph obtained from $K$ by removing edges in $b^-(G_A)$ and whole components of $G^b_A$ inside $K$; i.e. $K'$ consists of vertices and edges that belong both to $K$ and $G'$.

We apply the bound from Lemma~\ref{lem:potential_edge}:
$$\Phi(G') \ge \scc(G') + 1 + \Phi_E(G') - |V(G')|.$$
We decompose $\Phi_E(G') - |V(G')|$ into contributions from external edges and from the components of $\K$ reachable in $G_A$:
$$\Phi_E(G') - |V(G')| = \Phi_E(R_A) + \sum_{K \in \K_A} (\Phi_E(K') - V(K')).$$

Each avoidable component of $\K$ satisfies $|E(K')| \ge |V(K')|$, so we restrict our attention to unavoidable components.
By Lemma \ref{lem:unavoidable_GA}, each unavoidable component contributes $\Phi_E(K') - |V(K')| \ge -2$, and those with input and output in the same component contribute at least $-1$.

Let $t'$ be the number of unavoidable components whose input and output vertices lie in different components of $G^b_A$.
Observe that all these components lies on every directed cycle in $D(G, M_\pi) \setminus (A \setminus \set\pi)$ that contains $\pi$.
Therefore, the number of strongly connected components in $G'$ is $t' + 1$.
Thus, combining all contributions:
$$\Phi(G') \ge \scc(G') - 1 + \Phi_E(R_A) + \sum_{K \in \K_A} (|E(K')| - |V(K')|) \ge \Phi_E(R_A) - t.$$

We now prove that $\Phi_E(R_A) - t \ge \Phi_E(R'_B) - |\K'_B|$.
Since $G$ is strongly connected, every external edge is reachable in either $G_A$ or $G_B$, so $R'_B \subseteq R_A$.

We construct a set $Z$ of external edges that are reachable in both $G_A$ and $G_B$, containing one distinct edge for each component in $\K$ that is unavoidable in $G_A$ and reachable $G_B$.
For each such component $K$:
\begin{itemize}
    \item if $K$ lies in the ideal $I$, it has an incoming edge and this edge is reachable in both $G_A$ and $G_B$,
    \item similarly if $K$ lies in the complement of $I$, it has an outgoing edge.
\end{itemize}
Therefore, $|Z| = |\K'_B| - t$, and $R'_B \dotcup Z \subseteq R_A$, which implies:
$$\Phi_E(R_A) - t \ge \Phi_E(R'_B) - |\K'_B|.$$

Finally, if there exists a non-trivial component that is unreachable in $G_A$, or unavoidable with one input and one output in $G_A$, then by Lemma~\ref{lem:unavoidable_GA}, $G^b_A$ contains a non-trivial component excluded in $G'$.
Thus, $\Phi(G_A) \ge 2 \Phi(G') \ge 2(\Phi_E(R_A) - t)$, completing the proof.
\end{proof}

In Figure~\ref{fig:unavoidable}, we have $|R_A| = 9$ and $|R'_B| = 7$, so Lemma~\ref{lem:phi_GA} gives the lower bounds $\Phi(G_A) \ge 2(9 - 4) \ge 2(7 - 3).$

\begin{lemma} \label{lem:phi}
Assume that $A$ and $B$ split $\delta(I) \cup \set\pi$ for an ideal $I$ of $\hG$.
If $G_B$ contains a non-trivial and unreachable component $K_B \in \K$, and $G_A$ contains a non-trivial component $K_A \in \K$ such that $K_A$ is either unreachable in $G_A$ or unavoidable with only one input and one output vertex in $G_A$, then $A$ and $B$ charge the potential of $G$.
\end{lemma}
\begin{proof}
Using Lemmas~\ref{lem:phi_GA} and~\ref{lem:phi_GB}, we obtain
\begin{align*}
& \Phi(G_A) + \Phi(G_B) - \Phi(G) \\
&\ge 2(\Phi_E(R'_B) - |\K'_B|) + 2(\Phi_E(G) - \Phi_E(R'_B) - |V(G)| + |\K'_B| + 1) - \Phi(G) \\
&\ge |E(G)| - |V(G)| \ge \frac 13 |E(G)|.
\end{align*}
\end{proof}


\section{\texorpdfstring{Constructing sets $A$ and $B$}{Constructing sets A and B}}

In this section, we construct the sets $A$ and $B$ that charge the potential.
We assume that $G$ is a trimmed, strongly connected graph, and that $\pi = \sigma\tau$ is its $\M^+$-minimal edge with $b(G^-_\pi) = \emptyset$.
Furthermore, we assume that $G^+_\pi$ and $G^-_\pi$ do not charge the potential of $G$.
The structure of this section follows Algorithm~\ref{alg:AB}.

\begin{lemma} \label{lem:antichain}
Assume that there exist two non-trivial components $K_A$ and $K_B$ of $\K$ that are incomparable in $\hG$.
Let $I$ be the ideal of $\hG$ whose maximal elements are $K_A$ and $K_B$.
Let the outgoing edges of $I$ be split into $A$ and $B$ such that $B$ contains all edges outgoing from $K_B$.
Then $A$ and $B$ charge the potential of $G$.
\end{lemma}
\begin{proof}
Since $K_A$ is unreachable in $G_A$ and $K_B$ is unreachable in $G_B$, the result follows directly from Lemma~\ref{lem:phi}.
\end{proof}

\begin{lemma} \label{lem:two_edges}
Let $I$ be an ideal of $\hG$ such that every maximal element of $I$ has at least two outgoing edges, and every minimal element of the complement of $I$ has at least two incoming edges.
Then $\delta(I) \cup \set\pi$ can be split in time $\O{|E(G)|}$ into sets $A$ and $B$ such that every component of $\K$ is reachable in both $G_A$ and $G_B$.
\end{lemma}
\begin{proof}
Let $\K^1$ be the set of all maximal elements of $I$, and let $\K^2$ be the set of all minimal elements of the complement of $I$.
We construct the sets $A$ and $B$ such that both contain at least one outgoing edge from every component in $\K^1$ and at least one incoming edge to every component in $\K^2$.

These sets are constructed greedily, starting from empty sets $A$ and $B$.
We maintain the following invariant during the greedy algorithm: if a component in $\K^1 \cup \K^2$ has an incident edge in $A \cup B$, then it has at least one incident edge in both $A$ and $B$.

Each iteration constructs a sequence of components $K_1, \ldots, K_\ell \in \K$, and edges $e_1, \ldots, e_{\ell-1} \in \delta(I) \setminus (A \cup B)$ such that the end-vertices of $e_i$ belong to $K_i$ and $K_{i+1}$ for $i \in \set{1,\ldots,\ell-1}$, and components $K_2, \ldots, K_{\ell-1}$ have no incident edge in $A \cup B$.
Note that only $K_1$ and $K_\ell$ may coincide.
Such sequences of edges and components can be constructed greedily starting from a component in $\K^1 \cup \K^2$ that has no incident edge in $A \cup B$.
Edges of $e_1, \ldots, e_{\ell-1}$ with odd index are added to $A$ and edges with even index to $B$.
This approach ensures that the invariant is preserved.

Once every component of $\K^1 \cup \K^2$ has incident edges in both $A$ and $B$, the remaining edges of $\delta(I)$ can be assigned arbitrarily to $A$ and $B$.
This greedy algorithm processes in $\O{|E(G)|}$ time.

Now, consider any component $K$ of $I$.
It is reachable in $G_A$ because there exists a path from $\tau$ to some component $K' \in \K^1$ passing through $K$ in $G_A$.
Since $B$ contains an edge from $K'$ to a component $K''$ in the complement of $I$, and there is a path from $K''$ to $\sigma$ in the complement of $I$, it follows that every component of the complement is also reachable in $G_A$.
A similar argument shows that every component of the complement of $I$ is reachable in $G_A$ as well as every components of $\K$ is reachable in $G_B$.
\end{proof}

\begin{lemma} \label{lem:K1}
Assume that there are at least two edges from $\tau$ to $\kappa_1$.
Let $B$ contain exactly one such edge, and define $A = N_E(\tau) \setminus B$.
Then $A$ and $B$ charge the potential of $G$.
\end{lemma}
\begin{proof}
Since all components of $\K$ are reachable both in $G_A$ and $G_B$, Lemma \ref{lem:phi_GB} implies that $\Phi(G_B) = \Phi(G) - \Phi_E(B)$, and Lemma \ref{lem:phi_GA} implies that 
$$\Phi(G_A) \ge \Phi_E(R_A) - |\K| = \Phi_E(G) - \Phi_E(\K) - \Phi_E(A) - |\K|.$$
Combining the two expressions, we derive:
\begin{equation} \label{eq:K1_phi}
\Phi(G_A) + \Phi(G_B) - \Phi(G) \ge |E(G)| - |E(\K)| - |N_E(\tau)| - |\K|.
\end{equation}

The vertex $\tau$ has degree $|N_E(\tau)|$, and every other vertex in $V_\tau$ has degree at least three, so the number of edges in $G$ is at least:
\begin{equation} \label{eq:K1_edges}
|E(G)| \ge |N_E(\tau)| + \frac{3}{2} |V(G)| - 3.
\end{equation}

Moreover, since $\tau$ can be connected only to one partite set of $G$, it follows that:
\begin{equation} \label{eq:K1_vertices}
\frac{1}{2} |V(G)| \ge |N_E(\tau)|.
\end{equation}

We multiply inequality~\eqref{eq:K1_phi} by 1, \eqref{eq:K1_edges} by $3/4$, \eqref{eq:K1_vertices} by $1/4$, and inequality~\eqref{eq:phi_G_pi_basic} of Lemma~\ref{lem:phi_G_pi} by 1, and sum the resulting expressions:
$$\Phi(G_A) + \Phi(G_B) - \Phi(G) \ge \left( \frac{1}{4} - c \right)|E(G)| - \frac{1}{4} = \frac{1}{10} |E(G)| + \frac{1}{20} (3|E(G)| - 5),$$
which leads to a contradiction since $c = 0.1$. Note that the smallest trimmed graph, $K_{3,3}$, has 9 edges.
\end{proof}

After Lemma~\ref{lem:K1}, we may assume that $G$ has only one edge from $\tau$ to $\kappa_1$, and symmetrically, only one edge from $\kappa_\zeta$ to $\sigma$.
Since $\sigma$ has at least three neighbor, $\K$ contains at least two components, so $\zeta \ge 2$.

Since every vertex in $V(\kappa_1) \cap V_\sigma$, except the neighbor of $\tau$, has at least three neighbors in $\kappa_1$, it follows that $|E(\kappa_1)| \ge \frac32 |V(\kappa_1)| - 1$.
Using a similar inequality for $\kappa_\zeta$, we derive:
\begin{equation} \label{eq:K1}
|E(\kappa_1)| + |E(\kappa_\zeta)| \ge \frac32 \left(|V(\kappa_1)| + |V(\kappa_\zeta)|\right) - 2.
\end{equation}

\begin{lemma} \label{lem:one_one}
Assume that there exists a non-trivial component $K \in \K$ with exactly one incoming edge $uv$ and one outgoing edge $xy$.
Let $I$ be the ideal with maximal vertex $K$, and $B$ contains only the outgoing edge from $K$, and $A$ contains the remaining edges of $\delta(I) \cup \set\pi$.
Then $A$ and $B$ charge the potential of $G$.
\end{lemma}
\begin{proof}
From the construction it follows that $K$ is unavoidable in $G_A$ with only one input and one output vertex.
Furthermore, $K$ is unreachable in $G_B$.
By Lemma~\ref{lem:unavoidable_GA}, the conditions of Lemma~\ref{lem:phi} are satisfied, and the result follows.
\end{proof}

Since $\kappa_1$ has only one incoming edge, last lemma implies that $\kappa_1$ has at least two outgoing edges.
Next, we prove that $\kappa_2$ has at least two incoming edges.

\begin{lemma} \label{lem:K2}
Assume that component $\kappa_2$ has only one incoming edge $e$.
Let $I$ be the ideal containing $\tau$ and $\kappa_1$, and let $B$ contain only the edge $e$, while $A$ contains the remaining edges of $\delta(I) \cup \set\pi$.
Then $A$ and $B$ charge the potential of $G$.
\end{lemma}
\begin{proof}
The component $\kappa_2$ is unreachable in $G_B$ and non-trivial, since a trivial component requires at least two incoming edges.
Since $\kappa_1$ and $\kappa_2$ are comparable, the edge $e$ must be outgoing from $\kappa_1$.
The component $\kappa_1$ has only one input vertex, and by our construction of $A$, it also has only one output vertex in $G_A$.
By Lemma~\ref{lem:unavoidable_GA}, the conditions of Lemma~\ref{lem:phi} are satisfied, and the result follows.
\end{proof}

\begin{lemma} \label{lem:zeta_2}
It holds that $\zeta \ge 3$.
\end{lemma}
\begin{proof}
For contradiction, assume that $\zeta = 2$.
Then the set of components $\K$ consists only of $\kappa_1$ and $\kappa_2$, as there is no component incomparable with $\kappa_1$ and $\kappa_2$.

There are only the following external edges: one from $\tau$ to $\kappa_1$, one from $\kappa_2$ to $\sigma$, at most $|V(\kappa_2)|/2$ edges from $\tau$ to $\kappa_2$, at most $|V(\kappa_1)|/2$ edges from $\kappa_1$ to $\sigma$, and at most $|V(\kappa_1)||V(\kappa_2)|/4$ edges from $\kappa_1$ to $\kappa_2$.
Additionally, $G$ contains the edge $\pi$, so we obtain:
\begin{equation} \label{eq:zeta_edges}
|E(\K)| + 3 + \frac{1}{2} |V(\K)| + \frac{1}{4} |V(\kappa_1)||V(\kappa_2)| \ge |E(G)|.    
\end{equation}

From \eqref{eq:K1}, it follows that:
\begin{equation} \label{eq:zeta_internal}
|E(\K)| \ge \frac{3}{2} |V(\K)| - 2.
\end{equation}

Since both $\kappa_1$ and $\kappa_2$ contain at least 6 vertices, we obtain:
\begin{equation} \label{eq:zeta_vertices}
|V(\kappa_1)||V(\kappa_2)| \ge 36.
\end{equation}

From Lemma~\ref{lem:phi_G_pi}\eqref{eq:phi_G_pi_prod}, and the inequality \eqref{eq:K1} we derive:
\begin{align*}
c |E(G)| &\ge (|E(\kappa_1)| - |V(\kappa_1)| + 2)(|E(\kappa_2)| - |V(\kappa_2)| + 2) - 1 \\
&= (|E(\kappa_1)| - |V(\kappa_1)| + 1)(|E(\kappa_2)| - |V(\kappa_2)| + 1) \\
&\quad + |E(\kappa_1)| - |V(\kappa_1)| + |E(\kappa_2)| - |V(\kappa_2)| \\
&\ge \frac{1}{4} |V(\kappa_1)||V(\kappa_2)| + |E(\K)| - |V(\K)|.
\end{align*}

We now multiply the last inequality by 4, \eqref{eq:zeta_edges} by 1, \eqref{eq:zeta_internal} by 3, and \eqref{eq:zeta_vertices} by $3/4$, and sum these products. Hence, $4c|E(G)| \ge |E(G)| + 18$,
which leads to a contradiction since $0 < c \le 1/4$.
\end{proof}

\begin{lemma} \label{lem:AB}
Algorithm~\ref{alg:AB} finds sets $A$ and $B$ that charge the potential of $G$ in time $\O{|E(G)|}$.
\end{lemma}
\begin{proof}
Let $I_1$ be the ideal containing only $\tau$ and $\kappa_1$, and let $I_2$ be the ideal containing all elements of $\hG$ except $\sigma$ and $\kappa_\zeta$.
Since $\kappa_1$ has at least two outgoing edges, and $\kappa_2$ and all trivial components have at least two incoming edges, Lemma~\ref{lem:two_edges} implies that $\delta(I_1) \cup \set\pi$ can be split into $A$ and $B$ such that every component of $\K$ is reachable in both $G_A$ and $G_B$.
The same applies to $\delta(I_2) \cup \set\pi$.
Assume, for contradiction, that neither $I_1$ nor $I_2$ yields sets $A$ and $B$ that charge the potential of $G$.

We estimate $|E(G)|$ using the handshaking lemma: $2|E(G)| = \sum_{u \in V} \deg(u)$.
The vertices in $V(\kappa_1) \cup \set{\tau}$ have the following incident edges: $\delta(I_1)$, $\pi$, the internal edges $E(\kappa_1)$, and the edge from $\tau$ to $\kappa_1$.
Using \eqref{eq:K1}, we get:
$$\sum_{u \in V(\kappa_1) \cup \set{\tau}} \deg(u) = 2|E(\kappa_1)| + 3 + |\delta(I_1)| \ge 3|V(\kappa_1)| + 1 + |\delta(I_1)|.$$
A similar bound holds for $V(\kappa_\zeta) \cup \set{\sigma}$, and all remaining vertices have degree at least three. Thus,
\begin{equation}\label{eq:edges}
2|E(G)| \ge 3|V(G)| + |\delta(I_1)| + |\delta(I_2)| - 4.
\end{equation}

Since every vertex in $V_\tau \setminus \set{\tau}$ is incident with at least one edge in $E(\K)$, and $\kappa_1$ has at least two vertices with degree at least three in $E(\K)$, it follows that:
\begin{equation}\label{eq:edges_internal}
|E(\K)| \ge \frac{1}{2} |V(G)| + 3.
\end{equation}

\medskip
\textbf{Case: $\delta(I_1) \cap \delta(I_2) = \emptyset$.}

In this case, $E(G)$ contains all of $E(\K)$, $\delta(I_1)$, $\delta(I_2)$, the edge from $\pi$ to $\kappa_1$, the edge from $\kappa_\zeta$ to $\sigma$, and $\pi$ itself. So:
\begin{equation}\label{eq:delta_disjoint}
|E(G)| \ge |E(\K)| + |\delta(I_1)| + |\delta(I_2)| + 3.
\end{equation}

Every component of $\K$ is reachable in $G_A$, so by Lemma~\ref{lem:phi_GA}:
$$\Phi(G_A) \ge \Phi_E(E(G) \setminus (E(\K) \cup A)) - |\K|.$$

Since $A$ and $B$ do not charge the potential, Lemma \ref{lem:phi_GB} implies:
$$c|E(G)| > |E(G)| - |E(\K)| - |\delta(I_1)| - |\K| - 1.$$
A similar inequality holds for $I_2$, and combining both:
\begin{equation}\label{eq:phi_disjoint}
2 c|E(G)| > 2|E(G)| - 2|E(\K)| - |\delta(I_1)| - |\delta(I_2)| - 2|\K| - 2.
\end{equation}

Now multiply \eqref{eq:edges} by 3, \eqref{eq:edges_internal} by 2, \eqref{eq:delta_disjoint} by 2, \eqref{eq:phi_disjoint} by 5, and Lemma~\ref{lem:phi_G_pi}\eqref{eq:phi_G_pi_twice} by 5. Summing yields $15c|E(G)| \ge 2|E(G)| + 5$, which is a contradiction since $0 < c \le 2/15$.

\medskip
\textbf{Case: $\delta(I_1) \cap \delta(I_2) \neq \emptyset$.}

We now construct $A$ and $B$ such that every component in $\K$ is also avoidable in $G_A$.
Let $e_\delta \in \delta(I_1) \cap \delta(I_2)$, $e_\tau$ be an edge from $\tau$ to a component other than $\kappa_1$, and $e_\sigma$ an edge from a component other than $\kappa_\zeta$ to $\sigma$.
Note that $e_\delta$ may be equal to either $e_\tau$ or $e_\sigma$, but $e_\tau \ne e_\sigma$.
Then $\delta(I_1) \cup \set\pi$ can be split into $A$ and $B$ such that $B$ contains $e_\delta$, $e_\sigma$, and $e_\tau$, and every component in $\K$ is reachable in both $G_A$ and $G_B$
Furthermore, the three chosen edges ensure that all components in $\K$ are avoidable in $G_A$.

Since $\Phi(G_A) = \Phi_E(E(G) \setminus A) - |V(G)| + 2$, and similarly for $G_B$, it follows that:
$$c |E(G)| > \Phi(G_A) + \Phi(G_B) - \Phi(G) \ge |E(G)| - |\delta(I_1)| - |V(G)| + 1.$$
A similar inequality holds for $I_2$, hence:
\begin{equation}\label{eq:delta_intersect}
2 c |E(G)| > 2|E(G)| - |\delta(I_1)| - |\delta(I_2)| - 2|V(G)| + 2.
\end{equation}

From the inclusion-exclusion principle:
\begin{equation}\label{eq:inclusion_exclusion}
|E(G)| \ge |\delta(I_1)| + |\delta(I_2)| - |\delta(I_1) \cap \delta(I_2)| + |E(\K)| + 3.
\end{equation}

Since every edge in $\delta(I_1) \cap \delta(I_2)$ has one endpoint in $(V(\kappa_1) \cap V_\tau) \cup \set{\tau}$ and the other in $(V(\kappa_\zeta) \cap V_\sigma) \cup \set{\sigma}$, we get:
\begin{equation}\label{eq:delta_intersection}
\frac{1}{4} (|V(\kappa_1)| + 2)(|V(\kappa_\zeta)| + 2) \ge |\delta(I_1) \cap \delta(I_2)|.
\end{equation}

Finally, we multiply: \eqref{eq:edges} by 3, \eqref{eq:edges_internal} by 2, \eqref{eq:delta_intersect} by 5, \eqref{eq:inclusion_exclusion} by 2, \eqref{eq:delta_intersection} by 2, and Lemma~\ref{lem:phi_G_pi}\eqref{eq:phi_G_pi_prod} by 2. Summing yields $12 c |E(G)| \ge 2 |E(G)| + 8,$ which is a contradiction since $0 < c \le 1/6$.
\end{proof}


\section{Trimming}  \label{sec:trimming}

In this section, we show how to trim a graph $G$ to $G^T$ in time $\O{|E(G)| + \Phi(G^T) - \Phi(G)}$.
Since the edges in $b^-(G)$ can be removed and those in $b^+(G)$ can be merged into a single edge in time $\O{|E(G)|}$, and every non-trivial component of $G^b$ can be processed independently, we assume without loss of generality that $G$ is strongly connected to simplify the notation.

Recall that for two nodes $a$ and $b$ of the union-product circuit, we define $a \Square b$ as a new product node with children $a$ and $b$, and similarly, $a \cup b$ is a new union node.
By induction, we define $\Square_{i=1}^k a_i = a_1 \Square (\Square_{i=2}^k a_i)$ for nodes $a_1, \ldots, a_k$.
The node corresponding to an edge $e$ is denoted by $\eta_e$.

We begin by considering the case where $G$ is a cycle.

\begin{lemma} \label{lem:trim_cycle}
If $G$ is a cycle, then it can be trimmed into a single edge $e$ satisfying $\MM(e) = \MM(G)$ in time $\O{|V|}$ while creating $\O{|V|}$ new nodes.
\end{lemma}
\begin{proof}
The graph $G$ has exactly two perfect matchings, say $M_1$ and $M_2$.
Let $e$ be an edge whose node is given by $(\Square_{f \in M_1} \eta_f) \cup (\Square_{f \in M_2} \eta_f)$.
By definition, $\MM(e) = \MM(G)$.
The number of created nodes is $|V| - 1$, and the construction takes time $\O{|V|}$.
\end{proof}

Recall that the classical Hall’s marriage theorem \cite{hall1935representatives} states that a bipartite graph $G$ has a perfect matching if and only if $|N_V(W)| \ge |W|$ for every subset $W$ of vertices in one partite set of $G$.
Let $e$ be an edge with one endpoint in $N_V(W)$ and the other in $V \setminus W$.
If $|N_V(W)| = |W|$, then Hall’s theorem implies that no perfect matching in $G$ contains the edge $e$.
We use this observation in the following lemma.

\begin{lemma}\label{lem:trim_forest}
Let $W$ be the set of all vertices in one partite set of $G$ that have degree 2.
If the set of edges $N_E(W)$ contains a cycle, then $G$ is a cycle.
\end{lemma}
\begin{proof}
Let $C$ be a cycle in $N_E(W)$, and let $W' = V(C) \cap W$.
Then $|N_V(W')| = |W'|$.
By Hall’s theorem, every edge between $N_V(W')$ and $V \setminus W'$ belongs to $b^-(G)$.
Since $G$ is strongly connected, no such edge can exist, implying that $G$ contains only the vertices in $V(C)$.
Every vertex in $W$ has degree two, and so do the vertices in $N_V(W)$.
Therefore, $G$ is a collection of cycles.
As $G$ is connected, this collection must consist of a single cycle.
\end{proof}

Using Lemmas~\ref{lem:trim_cycle} and~\ref{lem:trim_forest}, we may assume that $N_E(W)$ forms a forest.
The first phase of trimming applies the following lemma to each component of $N_E(W)$.

\begin{figure}[tbp]
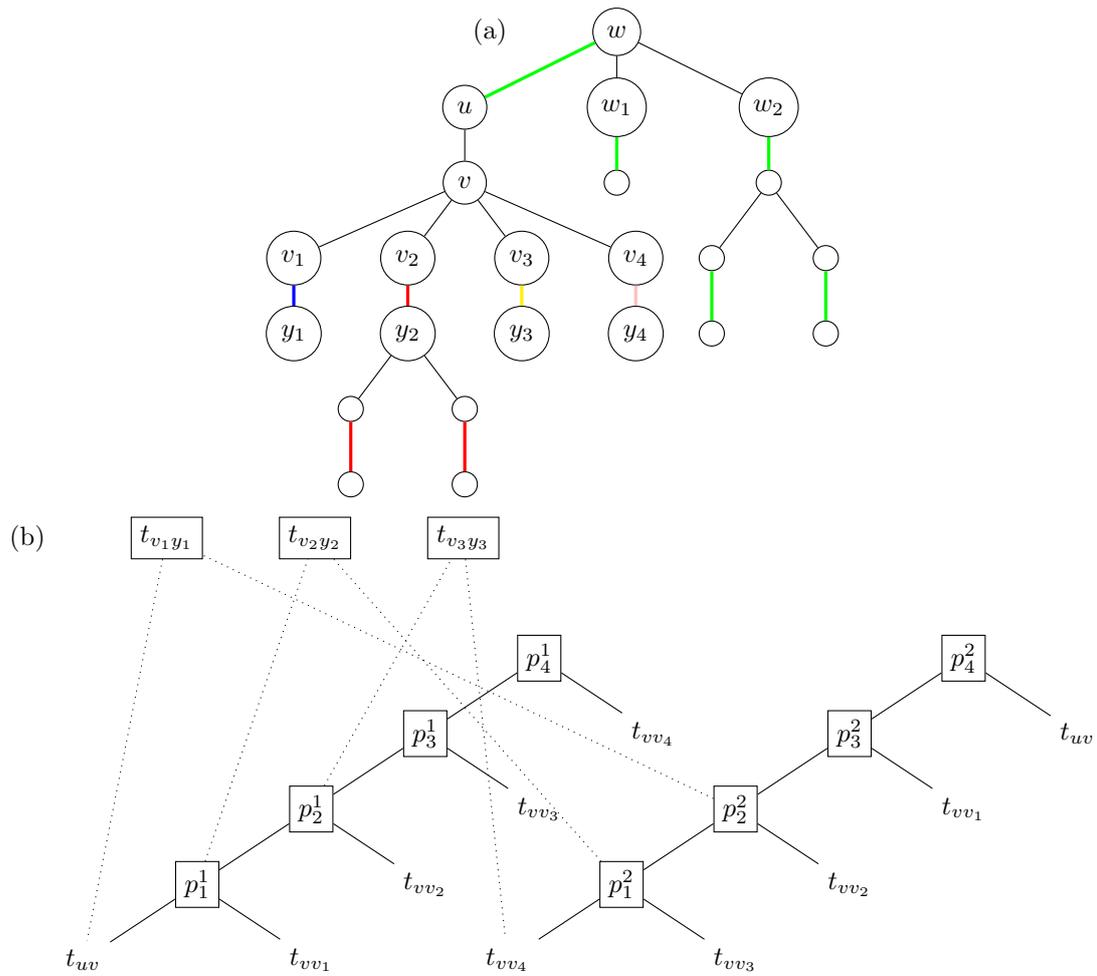

\centering
\include{trim_alg}
\caption{
(a) shows vertices $W$, their neighbors $N_V(W)$, and the edges $N_E(W)$, and illustrates how a matching $M_v$ is split into matchings $M_{uv}$ (green), $M_{vv_1}$ (blue), $M_{vv_2}$ (red), $M_{vv_3}$ (yellow), and $M_{vv_4}$ (pink).
(b) shows the construction of nodes $t_{v_1y_1}, t_{v_2y_2}$, and $t_{v_3y_3}$.
The dotted lines are used only for clarity.
}
\label{fig:trim_alg}
\end{figure}

\begin{lemma}\label{lem:trim_first_component}
Let $W$ contain some vertices with degree two from one partite set of a strongly connected graph $G$, such that $N_E(W)$ forms a tree.
Then the vertices of $W$ can be trimmed in time $\O{|N_E(N_V(W))|}$ while creating $\O{|W| \cdot |V|}$ new nodes.
\end{lemma}
\begin{proof}
Using Lemma~\ref{lem:trim_encoded}, we can trim the vertices of $W$ one by one by removing each $u \in W$ and its incident edges, and contracting the set $N_V(W)$ into a single vertex $z$ with neighbors $N_V(N_V(W)) \setminus W$.
If two or more vertices in $N_V(W)$ share a common neighbor $v \in V \setminus W$, the resulting graph must contain only one edge between $z$ and $v$.
This contraction can be performed in $\O{|N_E(N_V(W))|}$ time.
The main challenge lies in updating the nodes of the union-product circuit for edges incident to $z$ efficiently.

Let $M_v$ be the unique maximum matching in $N_E(W)$ that avoids vertex $v$, for every $v \in N_V(W)$.
Observe that $|M_v| = |W|$, and any perfect matching in $G$ that includes an edge $vy$ with $y \notin W$ must also include all edges in $M_v$.
Similarly to Lemma~\ref{lem:trim_encoded}, the node $\eta_{yz}$ for the new edge $zy$ is $\eta_{vy} \Square t_v$, where $t_v = \Square_{e \in M_v} \eta_e$, constructed as follows.

For an edge $uv$ with $u \in W$ and $v \in N_V(W)$, let $F_{uv}$ be the component of $N_E(W) \setminus \set{uv}$ containing $u$, and let $M_{uv}$ be the unique maximum matching in $F_{uv}$.
Then $M_v = \bigcup_{a \in W \cap N_V(v)} M_{av}$; see Figure~\ref{fig:trim_alg}(a).
We first construct nodes $t_{uv} = \Square_{e \in M_{uv}} \eta_e$ for every $uv \in N_E(W)$, and then $t_v = \Square_{a \in W \cap N_V(v)} t_{av}$.
A naïve construction of all $t_e$ for $e \in N_E(W)$ would take $\O{|W|^2}$ time in the worst case (e.g., if $N_E(W)$ is a star), so we use a more efficient approach.

Let us choose one vertex of $N_V(W)$ as the root $r$ of the tree $N_E(W)$.
Classify the edges of $N_E(W)$ into \emph{even} and \emph{odd} based on the parity of the distance from the root $r$; odd edges include those incident to the root $r$ and even edges include those incident to leaves.

We construct nodes $t_{uw}$ for odd edges $uw$ of $N_E(W)$ recursively from leaves to root.
Since $uw$ is odd, $w$ is closer to the root than $u$.
Let $v_1, \ldots, v_k$ be all neighbors of $v$ in $W$ except $u$, and let $y_1, \ldots, y_k$ be the other neighbors of $v_1, \ldots, v_k$, respectively.
Then $t_{uw}$ is constructed as $t_{uw} = \eta_{uv} \Square (\Square_{i=1}^k t_{vv_i})$, since $M_{uw} = \set{uv} \cup \bigcup_{i=1}^k M_{vv_i}$.

For even edges, we process from root to leaves.
We use the same notation and construct nodes $t_{v_1y_1}, \ldots, t_{v_ky_k}$ simultaneously.
All nodes $t_{vv_i}$ and $t_{uv}$ are already constructed.
We first build two chains of nodes $p^1_i$ and $p^2_i$ for $i = 1, \ldots, k$.
In $p^1_i$, the left child is $p^1_{i-1}$ and the right is $t_{vv_i}$ (with $p^1_1$ having left child $t_{uv}$).
In $p^2_i$, the order is reversed; see Figure \ref{fig:trim_alg}(b).
Then:
\begin{itemize}
    \item $t_{v_1y_1} = t_{uv} \Square p^2_{k-2}$,
    \item $t_{v_{k-1}y_{k-1}} = p^1_{k-2} \Square t_{vv_k}$,
    \item $t_{v_ky_k} = p^1_{k-1}$, and
    \item for $i = 2, \ldots, k-1$: $t_{v_iy_i} = p^1_{i-1} \Square p^2_{k-i-1}$.
\end{itemize}

The time complexity to construct all nodes $t_v$ is $\O{|W|}$.
Processing the remaining edges in $N_E(N_V(W)) \setminus N_E(W)$ takes $\O{|N_E(N_V(W))|}$ time.
The total number of created nodes is $\O{|W| + |N_E(N_V(W))|} = \O{|W||V|}$.
\end{proof}

\begin{lemma} \label{lem:trim_first}
The first phase of trimming has time complexity $\O{|E|}$.
\end{lemma}
\begin{proof}
Let $W$ be the set of all vertices from one partite of $G$ with degree two.
We partition $W$ into $W_1, \ldots, W_k$ such that $N_E(W_1), \ldots, N_E(W_k)$ form the connected components of $N_E(W)$.
We apply the previous lemma to each set $W_1, \ldots, W_k$ independently.
The total number of created nodes is $\O{|W||V|}$.
Observe that the sets of edges $N_E(N_V(W_1)), \ldots, N_E(N_V(W_k))$ are pairwise disjoint, so the time complexity is $\O{|E|}$.
We then process the vertices of degree two in the other partite of $G$ analogously, which completes the proof.
\end{proof}

\begin{lemma}\label{lem:trim_potential}
After the first phase, every vertex of degree two has an incident edge with potential at least two.
\end{lemma}
\begin{proof}
Let the partites of vertices in $G$ be denoted by $V_1$ and $V_2$.
The first phase applies Lemma \ref{lem:trim_first_component} to all vertices of degree two in $V_1$, followed by those in $V_2$.
Since applying Lemma \ref{lem:trim_first_component} yields the same result as trimming vertices of degree two one by one, consider the trimming of a vertex $u \in V_1$ with neighbors $v$ and $w$.

Consider trimming a vertex $u \in V_1$ with only neighbors $v$ and $w$ that causes a vertex $y$ to become degree two.
If $y \in V_2$, then $y$ is subsequently removed when vertices of degree two in $V_2$ are trimmed.
So, suppose $y \in V_1$.
The only way the degree of $y$ could decrease is if it was connected to both $v$ and $w$.
In this case, contracting $v$ and $w$ into a new vertex $z$ creates parallel edges between $y$ and $z$.
Joining them forms an edge $yz$ with potential
$$\Phi(yz) = \Phi(yv)\Phi(uw) + \Phi(yw)\Phi(uv) \ge 2.$$
Therefore, after processing vertices in $V_1$, every vertex of $V_1$ with degree two has an incident edge with potential at least two.

Next, consider trimming a vertex $u \in V_2$ that causes a vertex $y$ to become degree two.
If $y \in V_2$, the same analysis applies as above.
If $y \in V_1$, then $y$ must be the vertex $z$ created by contracting $v$ and $w$, since $v, w,$ and $y$ are the only vertices from $V_1$ involved in this trimming.
Let $a$ and $b$ be the neighbors of $z$.
The only possible neighbors of $v$ and $w$ are $u$, $a$ and $b$.
If any edge incident to $v$ or $w$ had potential at least two, then $z$ inherits an incident edge with potential at least two.
Since $v, w \in V_1$, they cannot have degree two, so they both have all $u$, $a$, and $b$ as neighbors.
The contraction creates parallel edges which are joined into a single edge with potential at least two.
\end{proof}

Note that the last lemma also holds during the second phase, however, a stronger statement is required there.

For the second phase, consider a vertex $u \in V(G)$ of degree two with neighbors $v$ and $w$, which are contracted into a new vertex denoted $v^2$.
By Lemma \ref{lem:trim_potential}, the edge $uv$ has potential at least two (without lost of generality).
By Lemma \ref{lem:trim_encoded}, trimming $u$ has time complexity $\O{\deg(v) + \deg(w)}$.
This trimming increases the potential of $G$ by:
$$(\Phi(uv)-1) \sum_{e \in N_E(w) \setminus \set{uw}} \Phi(e) - \Phi(uv) + (\Phi(uw)-1) \sum_{e \in N_E(v) \setminus \set{uv}} \Phi(e) - \Phi(uw) + 2.$$
Since $\Phi(uv) \ge 2$, the increase in potential is at least $\Omega(\deg w)$, which pays for the time required to update the nodes of all edges in $N_E(w) \setminus \set{uw}$.
For every vertex $y \in N_V(w) \setminus \set{u}$, we can test whether $vy \in E$ using the adjacency matrix of $G$, and, if so, merge the resulting parallel edges into a single edge with a new node.

However, this increase in potential is insufficient to pay the time updating the nodes of edges incident to $v$ if $\Phi(uw) = 1$ and $\deg(v) \gg \deg(w)$.
Thus, we postpone updating these nodes.
To manage this, the vertex $v^2$ keeps two separate groups of neighbors: one for edges originally connected to $w$ (including merged edges), and another for edges originally connected to $v$.
Additionally, $v$ stores $\eta_{uw}$ constructed by Lemma \ref{lem:trim_first_component}, which is needed later to update edges in the second group.
The edges $v^2y$ with nodes $\eta_{v^2y}$ of in the second group are called \emph{delayed}, and will be updated to $\eta_{v^2y} \Square \eta_{uw}$ later.

Suppose the original graph $G$ has a vertex $u^2$ with only neighbors $v$, $w$, and $w^2$.
After contracting $v$ and $w$, $u^2$ becomes degree two, and the process repeats.
Again, we delay the updates of nodes for edges incident to $v^2$, creating another group of neighbors in $v^3$, the result of contracting $v^2$ and $w^2$.
After $l-3$ iterations, vertex $v^l$ has its neighbors split into $l$ groups, each with a node that must be combined with the node of each edge in its group.
Each iteration requires applying the $\Square$ operation between the node of each group and the node $\eta_{v^l w^l}$, requiring additional time $\O{l}$.
To pay this time, observe:
$$\Phi(v^2 u^2) = \Phi(uv) \Phi(u^2 w) + \Phi(uw) \Phi(v u^2) \ge 3,$$
and by induction, $\Phi(v^l u^l) \ge l+1$, so the increase in the potential is $\Omega(l + \deg(w^l))$.

Further details are provided in the following lemma.
Note that every group of edges has a node except one.
To avoid exceptions, we assume the missing node is empty, and the product of a node with an empty node is defined to be the node itself.

\begin{lemma} \label{lem:trim}
A strongly connected graph $G$ can be trimmed into a graph $G^T$ such that $\MM(G) = \MM(G^T)$ in time $\O{|E(G)| + \Phi(G^T) - \Phi(G)}$ while creating 
$$\O{|V(G)| \cdot (|V(G)| - |V(G^T)|) + |E(G)| - |E(G^T)|}$$
new nodes in the union-product circuit.
\end{lemma}
\begin{proof}
We assume Lemma~\ref{lem:trim_first} has already been applied, so the graph $G$ satisfies Lemma~\ref{lem:trim_potential}.
During the second phase, the following data structures are maintained and the associated invariants are preserved:

\begin{enumerate}
    \item Each vertex $v$ has its incident edges partitioned into $l_v$ groups, and each group has an associated node in the union-product circuit.
    Exactly one of these nodes is empty.
    \item For an edge $uv$ with a delayed node $\eta_{uv}$ belonging to group $i$ in $u$ and group $j$ in $v$ (with nodes $\eta^i_u$ and $\eta^j_v$ respectively), the actual node for $uv$ is $\eta_{uv} \square \eta^i_u \square \eta^j_v$.
    \item For every vertex $u$ with only neighbors $v$ and $w$, it holds that $\Phi(uw) \ge 2$ or $\Phi(uv) \ge \max\set{2, l_v}$; and symmetrically, $\Phi(uv) \ge 2$ or $\Phi(uw) \ge \max\set{2, l_w}$.
\end{enumerate}

Initially, each vertex has all incident edges in a single group with an empty node.
Combined with Lemma~\ref{lem:trim_potential}, this ensures the invariants hold at the beginning.

\medskip
\textbf{Clearing operations.}
Clearing a vertex $v$ involves updating each edge $uv$ with a delayed node in group $i$ by replacing $\eta_{uv}$ with $\eta_{uv} \square \eta^i_u$ and moving all such edges to the group in $v$ with the empty node.
Clearing an edge $uv$ involves replacing its delayed node with $\eta_{uv} \square \eta^i_u \square \eta^j_v$ and moving it to the group with the empty node in both $u$ and $v$ (where $i$ and $j$ are the respective groups).
In both cases, groups with no edges are deleted.
These operations maintain the invariants, and their time complexities are $\O{\deg v}$ for clearing a vertex and $\O{1}$ for clearing an edge.

\medskip
\textbf{Second phase trimming algorithm.}
The algorithm proceeds in iterations until no vertex of degree two remains. Each iteration follows one of two cases:

\medskip
\textbf{Case 1:} A vertex $u$ of degree two has both incident edges with potential at least two.
In this case, we clear the vertices $u$, $v$, and $w$, and then trim $u$.

\medskip
\textbf{Case 2:} Every vertex with degree two has exactly one incident edge with potential one.

Let $v$ be a vertex, and let $u_1, \ldots, u_k$ be all its neighbors of degree two such that edges $u_1v, \ldots, u_kv$ have potential at least two.
Then, for each $i$, the other edge incident to $u_i$ is $u_iw_i$ with potential one.
We aim to remove all vertices $u_1, \ldots, u_k$ and their incident edges, and contract $v, w_1, \ldots, w_k$ into a new vertex $z$.

We first clear all vertices $u_1, \ldots, u_k$ and $w_1, \ldots, w_k$.
Following the method in Lemma~\ref{lem:trim_first_component}, we construct nodes $t_v, t_{w_1}, \ldots, t_{w_k}$ for the set $W = \{u_1, \ldots, u_k\}$.
Each node $\eta^j_v$ for group $j$ in $v$ is updated to $\eta^j_v \square t_v$, and a new group with an empty node is created.
Each edge $w_iy$ with node $\eta_{w_iy}$ is replaced by an edge $zy$ with node $\eta_{w_iy} \square t_{w_i}$ and moved into the group with the empty node in $z$.
Parallel edges in $z$ are detected using the adjacency matrix and merged accordingly.
Note that $z$ gains at most one more group than $v$ had.

This algorithm ensures that Invariants 1 and 2 are preserved.
We now verify Invariant 3.
Since edge potentials do not decrease and only $z$ may gain an additional group, it suffices to check Invariant 3 for vertices whose degree was reduced to two in the last iteration.
There are two such cases:

\medskip
\textbf{Case A:} Vertex $z$ has degree two, with neighbors $a$ and $b$, and $\Phi(za) = 1$.
Then $a$ cannot be adjacent to both $v$ and $w_i$, so without loss of generality, suppose $a$ was only adjacent to $v$.
This implies that $w_i$ has only $u_i$ and $b$ as neighbors.
Then Invariant 3 applied to $w_i$ implies $\Phi(zb) \ge \Phi(w_ib) \ge l_b$ as required.

\medskip
\textbf{Case B:} A vertex $y$ has only one neighbor $x$ outside of $v, w_1, \ldots, w_k$, and at least two neighbors among $v, w_1, \ldots, w_k$.
In this case, $y$ is connected to some $w_i$.
From Invariant 3, we have $\Phi(vu_i) \ge l_v$, which implies $\Phi(zw_i) \ge l_v + 1$, as required.

\medskip
\textbf{Time complexity analysis.}
The first phase and the final clearing of all vertices after the second phase take time $\O{|E(G)|}$, by Lemma~\ref{lem:trim_first}.
We prove that one iteration of the second phase that trims $G$ into $G'$ has time complexity $\O{\Phi(G') - \Phi(G) + |V(G)| - |V(G')|}$.
Then, using the telescopic sum over all iterations, the time complexity stated in this lemma follows.

In Case 1, then the potential increases by at least $\deg(v) + \deg(w) - 2$, and trimming $u$ takes $\O{\deg(v) + \deg(w)}$ time.
So, the time complexity is $\O{\Phi(G') - \Phi(G) + |V(G)| - |V(G')|}$ as required.

In Case 2, each edge $yw_i$ (with $y \ne u_i$) has its potential increased by $\Phi(yw_i)(\Phi(vu_i)-1)$.
Additionally, the edges $vu_i$ and $u_iw_i$, as well as the vertices $u_i$ and $w_i$, are removed.
So the total potential increase is:
\begin{align*}
& k - \sum_{i=1}^k \Phi(vu_i) + \sum_{i=1}^k (\Phi(vu_i)-1) \sum_{y \in N_V(w_i) \setminus \set{u_i}} \Phi(yw_i) \\
& = \sum_{i=1}^k (\Phi(vu_i)-1) \left( \sum_{y \in N_V(w_i) \setminus \set{u_i}} \Phi(yw_i) - 1 \right) \\
& \ge \max \set{1, l_v - 1} \cdot \sum_{i=1}^k \max \set{1, \deg(w_i) - 2} \\
& \ge \frac{1}{2} \left( k(l_v - 1) + \sum_{i=1}^k (\deg(w_i) - 2) \right) \\
& \ge \frac{1}{2} \left( l_v + \sum_{i=1}^k \deg(w_i) - 3k \right)
\end{align*}
where the inequality $\sum_{y \in N_E(w_i) \setminus \set{u_i}} \Phi(yw_i) \ge \max\set{2, \deg(w_i)-1}$ follows from Lemma~\ref{lem:trim_potential} applied to $w_i$.

This potential increase pays for the trimming time $\O{l_v + \sum_{i=1}^k \deg(w_i)}$, up to the additive term $\frac{3}{2}k = \O{|V(G)| - |V(G')|}$, similarly as in the Case 1.

\medskip
\textbf{Space complexity analysis.}
For each trimmed vertex, we create $\O{|V(G)|}$ new nodes.
For each removed parallel edge, we create only one new node.
\end{proof}


\section{Visiting trees of union-product circuits}

\begin{lemma} \label{lem:visit_matching}
Let $u$ be a node in the union-product circuit.
Then, there exists a bijection between the set of visiting trees of $u$ and the set of matchings in $\Upsilon(u)$, such that for each visiting tree $S$ and its corresponding matching $M \in \Upsilon(u)$, the set of edges stored in the leaves of $S$ is precisely $M$.
\end{lemma}
\begin{proof}
We prove the lemma by induction. 
If $u$ is a leaf node containing an edge $e$, then there exists exactly one visiting tree, whose only leaf contains the edge $e$.
In this case, $\Upsilon(u) = \set{\set{e}}$, and the correspondence is immediate.

Suppose now that $u$ is a union node with children $l$ and $r$.
By definition, each visiting tree rooted at $u$ contains either a visiting tree of $l$ or a visiting tree of $r$.
Moreover, the set of matchings satisfies $\Upsilon(u) = \Upsilon(l) \dotcup \Upsilon(r)$. Hence, there is a bijection between visiting trees of $u$ and matchings in $\Upsilon(u)$, established via the bijections for $l$ and $r$.

Now, suppose $u$ is a product node.
Then, for every visiting tree of $u$, its set of leaves is the union of the leaves of a visiting tree in $l$ and a visiting tree in $r$.
Similarly, each matching in $\Upsilon(u) = \Upsilon(l) \Square \Upsilon(r)$ is the union of a matching from $\Upsilon(l)$ and one from $\Upsilon(r)$.
By the induction hypothesis, the correspondence between visiting trees and matchings for $l$ and $r$ is bijective, which extends naturally to $u$.
\end{proof}

\begin{algorithm}[ht]
\caption{Enumerating all visiting trees rooted a given node $r$}
\label{alg:visit}
\myproc{\Visit{$r$}}{
    Initialize auxiliary nodes and both stacks\;
    \While{the ready stack is non-empty}{
        $(u,v,d)$ := pop from the ready stack\;
        \eIf{$v$ is the last node in $P^d_u$}{
            \uIf{$d = \xright$ and the left child of $u$ has potential at least two}{
                push $(u,\xright)$ to the waiting stack\;
            }
            \uElseIf{$d = \xleft$ and the right child of $u$ has potential at least two}{
                pop from the waiting stack \tcp{Removes $(u,\xright)$}
            }
        }{
            $w$ := the next node of $P^d_u$ after $v$\;
            \Ready{$u,w,d$}\;
            \While{the waiting stack is non-empty}{
                $(u,d)$ := pop from the waiting stack\;
                $v$ := the first node of $P^d_u$\;
                \Ready{$u,v,d$}\;
            }
            Visit the perfect matching in the visiting tree rooted at $\visit_{a_3,\xright}$\;
        }
    }
}

\myproc{\Ready{$u,v,d$}}{
    $\visit_{u,d} := v$\;
    Push $(u, v, d)$ to the ready stack\;
    \If{$\Phi(v) \ge 2$}{
        Push $(\xskip_v, \xright)$ and $(\xskip_v, \xleft)$ to the waiting stack\;
    }
}
\end{algorithm}

\begin{lemma} \label{lem:visit_tree}
Algorithm~\ref{alg:visit} enumerates every visiting tree of a given node $r$ in the union-product circuit exactly once.
\end{lemma}
\begin{proof}
To enumerate all visiting trees, each node $u$ maintains five auxiliary pointers. 
First, the pointers $\up_{u,\xleft}$ and $\up_{u,\xright}$ represent the nodes constructed through trimming.
Second, the pointers $\visit_{u,\xleft}$ and $\visit_{u,\xright}$ represent the visiting tree used to construct a perfect matching during execution of the function \Visit.
Third, the pointer $\xskip_u$ is defined later in this proof.

Note that only product nodes and leaves appear in visiting trees, since a union node always selects exactly one of its children.
Define the set $P_u$ recursively as follows: if $u$ is a product node or a leaf, then $P_u = \{u\}$; otherwise, if $u$ is a union node with children $l$ and $r$, then $P_u = P_l \cup P_r$.
For a product node $u$ with children $l$ and $r$, define $P^{\xleft}_u = P_l$ and $P^{\xright}_u = P_r$.
For each product node $u$, the pointers $\visit_{u,\xleft}$ and $\visit_{u,\xright}$ must point to elements in $P^{\xleft}_u$ and $P^{\xright}_u$, respectively.
Although the sets $P^{\xleft}_u$ and $P^{\xright}_u$ are defined to simplify the description of the function \Visit, they need not be stored explicitly, since a depth-first search (DFS) traversal can enumerate these sets with constant delay.

Analogous to standard trees, a node $v$ is a descendant of a node $u$ if the union-product circuit has a directed path from $u$ to $v$.
Furthermore, $u$ is a predecessor of $v$ if $v$ is a successor or $u$.

If a node $u$ has potential one, then it is a leaf or a product node and none of its descendants is a union node.
In this case, the function \Visit does not need to traverse $u$'s descendants, which is essential for achieving the desired time complexity.
Consequently, in nodes with potential one, the pointers $\visit_{u,\xleft}$ and $\visit_{u,\xright}$ are set equal to $\up_{u,\xleft}$ and $\up_{u,\xright}$ at the time of creation.

In addition, if both children of a product node $u$ are themselves product nodes and only one has potential at least two, the pointer $\xskip_u$ points to the parent of the nearest union node among $u$'s descendants.
This pointer is initialized during node creation as follows:
\begin{itemize}
    \item Nodes with potential one, union nodes, and leaves do not have a $\xskip_u$ pointer.
    \item If a product node $u$ has at least one child that is a union node, or if both children have potential at least two, then $\xskip_u := u$. Such nodes are called \emph{free nodes}.
    \item If both children of $u$ are product nodes and one has potential one, then $\xskip_u$ is set to $\xskip_v$, where $v$ is the child with potential at least two.
    Moreover, during creation, the pointers $\visit_{u,\xleft}$ and $\visit_{u,\xright}$ are set to the respective children of $u$.
\end{itemize}
Therefore, function \Visit must update $\visit_{u,\xleft}$ and $\visit_{u,\xright}$ only for free nodes.

The order in which visiting trees rooted at a product node $u$ are enumerated satisfies the following recursive rules, applied in order:
\begin{enumerate}
    \item Traverse the set $P^{\xleft}_u$ from left to right.
    \item For a fixed left pointer $\visit_{u,\xleft}$, recursively enumerate all its visiting trees.
    \item For each visiting tree of the left child, traverse $P^{\xright}_u$ from left to right.
    \item For each right pointer $\visit_{u,\xright}$, recursively enumerate all its visiting trees.
\end{enumerate}
In other words, for each visiting tree of the left child, we enumerate all visiting trees of the right child before moving to the next visiting tree of the left child.
For example, the matchings in Figure~\ref{fig:trimming} are visited in the order:
$$
\{uv, wv_1, u'v', w'v'_1\}, \quad 
\{uv, wv_1, v'v'_1, u'w'\}, \quad 
\{vv_1, uw, u'v', w'v'_1\}, \quad 
\{vv_1, uw, v'v'_1, u'w'\}
$$
and those in Figure~\ref{fig:circuit} are visited in the order:
$$\{h, e, a\}, \quad \{h, c, b\}, \quad \{a, f, g\}, \quad \{b, d, g\}.$$

Since the function \Visit dynamically modifies $\visit$ pointers, we distinguish their states:
We say that a pointer $\visit_{u,d}$ (where $d \in \{\xleft, \xright\}$) is \emph{ready} if it points to a node in $P^d_u$.
It is said to be \emph{waiting} if it is not ready, and:
\begin{itemize}
    \item some $\visit$ pointer in an ancestor of $u$ points to $u$, or
    \item some $\visit$ pointer points to a product node $v$ with $\xskip_v = u$.
\end{itemize}
At the moment a visiting tree is enumerated, all its $\visit$ pointers must be in the ready state.

We maintain the \emph{visit invariant} that 
\begin{itemize}
    \item both visit pointers in all free ancestors of any waiting or ready pointer are ready, and
    \item both visit pointers in all free descendants of any waiting pointer are neither waiting nor ready.
\end{itemize}

The function \Visit employs two stacks: the \emph{ready} stack and the \emph{waiting} stack.
The ready stack contains all triples $(u, v, d)$ such that $u$ is a free node, $v \in P^d_u$, and $\visit_{u,d} = v$.
The waiting stack contains pairs $(u, d)$ where $\visit_{u,d}$ is currently waiting.

The auxiliary function \Ready{$u,v,d$} transitions a pointer $\visit_{u,d}$ from the waiting to the ready state by setting it to $v \in P^d_u$ and pushing the corresponding triple to the ready stack.
If the potential of $v$ is at least two, then $(\xskip_v,\xright)$ and $(\xskip_v,\xleft)$ are pushed onto the waiting stack (in this order) to preserve the left-to-right enumeration order.
This ensures that the visit invariant remains satisfied.

The main loop of \Visit consists of two nested loops.
The outer loop processes the ready stack until it reaches a triple $(u,v,d)$ where $v$ is not the last node in $P^d_u$.
As ready triples are removed, the corresponding pointers become waiting.
If $d = \xright$ and the left child of $u$ has potential at least two, then the next ready triple is $(u,w,\xleft)$ for some $w \in P^{\xleft}_u$.
Hence, we push $(u,\xright)$ to the waiting stack to delay updating $\visit_{u,\xright}$ until after updating $\visit_{u,\xleft}$.
Conversely, if $d = \xleft$ and the right child has potential at least two, then the next triple on the ready stack points to $u$ via its predecessors, so we remove the pair $(u,\xright)$ from the top of the waiting stack (which had been added previously).

Once a triple $(u,v,d)$ is removed and $v$ has a successor $w$ in $P^d_u$, we set $\visit_{u,d} := w$ via \Ready{$u,w,d$}.
Then the inner loop processes the waiting stack until it is empty.
Each waiting pair $(u,d)$ is resolved by calling \Ready{$u,v,d$} where $v$ is the first node in $P^d_u$, ensuring that all descendants are properly initialized.

To initialize the process, we construct a small auxiliary structure above $r$:
\begin{itemize}
    \item A product node $a_1$ with no children,
    \item A union node $a_2$ with left child $a_1$ and right child $r$,
    \item A product node $a_3$ with left child $a_2$ and no right child,
    \item Set $\visit_{a_3,\xleft} := a_1$,
    \item Initialize the ready stack with the triple $(a_3, a_1, \xleft)$,
    \item The waiting stack is initially empty.
\end{itemize}

The first step of function \Visit is to remove this triple and invoke \Ready{$a_3, T, \xright$}, where $T$ is the first node in $P^{\xright}_{a_3}$.
This initiates enumeration of the leftmost visiting tree rooted at $r$.

Thus, the function \Visit preserves the visit invariant, visits each visiting tree exactly once, and proceeds according to the specified enumeration order.
\end{proof}

\begin{lemma} \label{lem:visit_time}
The time complexity of Algorithm~\ref{alg:visit}, when called on a node $r$, is $\mathcal{O}(\Phi(r))$.
\end{lemma}
\begin{proof}
The total time complexity of Algorithm~\ref{alg:visit} is determined by the number of iterations of the inner and outer loops.

For each free node $u$, let $D_u$ denote the number of iterations in which:
\begin{itemize}
    \item a pair $(v, d)$ is popped from the waiting stack, or
    \item a triple $(v, w, d)$ is popped from the ready stack,
\end{itemize}
where $v$ is $u$ or a descendant of $u$.
Note that the node $u$ may be entered multiple times from its parent during enumeration.
Thus, the value $D_u$ must be accounted for multiple times in its parent, since $D_u$ reflects only one traversal into $u$.

We proceed by induction from leaves to the root $r$ and prove that:
$$D_u \le 6 \Phi(u) - 9.$$

Let $u$ be a free node with left and right children $l$ and $r$, respectively.

\paragraph{The base of induction.}
Assume that $u$ has no free descendants.
That is, both $P^{\xleft}_u$ and $P^{\xright}_u$ consist solely of product nodes with potential one, and at least one of them contains at least two nodes.
We consider two cases:

\smallskip
\textit{Case 1:} One of the children (say, $l$) is a product node, and the other ($r$) is a union node.

In this case, $P^{\xleft}_u = \{l\}$, and $|P^{\xright}_u| \ge 2$.
Upon entering $u$, the inner loop executes once for the pair $(u,\xright)$, and the outer loop executes once for each triple $(u, v, \xright)$ with $v \in P^{\xright}_u$.
Thus:
\[
D_u = |P^{\xright}_u| + 1 = \Phi(u) + 1 \le 6 \Phi(u) - 9, \quad \text{since } \Phi(u) = |P^{\xright}_u| \ge 2.
\]

The case where $r$ is a product node is symmetric.

\smallskip
\textit{Case 2:} Both $l$ and $r$ are union nodes. The function performs the following loop iterations:
\begin{itemize}
    \item One iteration of the inner loop for the pair $(u,\xleft)$ to set $\visit_{u,\xleft}$ to the first node $v^{\xleft}_1 \in P^{\xleft}_u$.
    \item One iteration of the inner loop for $(u,\xright)$ to set $\visit_{u,\xright}$ to the first node $v^{\xright}_1 \in P^{\xright}_u$.
    \item For each $v^{\xright}_i \in P^{\xright}_u$, an outer loop iteration to update $\visit_{u,\xright}$.
    \item After processing all right children, an outer loop iteration updates $\visit_{u,\xleft}$ to $v^{\xleft}_2$, and the process repeats.
\end{itemize}
Therefore, the number of iterations is:
$$D_u = 1 + |P^{\xleft}_u| \cdot (|P^{\xright}_u| + 1) = 1 + \Phi(l) + \Phi(l)\Phi(r) \le 6\Phi(u) - 9 + \left(10 + \Phi(l)(1 - 5\Phi(r))\right).$$
Since both $\Phi(l), \Phi(r) \ge 2$, the bracketed term is non-positive, ensuring the bound holds.

\paragraph{Inductive Step.}
Assume now that $u$ has free descendants. We consider two cases.

\smallskip
\textit{Case 1:} The left child $l$ has potential one, and the right child $r$ has potential at least two.

Since $P^{\xleft}_u = \{l\}$ and $|P^{\xright}_u| \ge 2$, the algorithm performs:
\begin{itemize}
    \item One iteration of the inner loop for $(u,\xright)$,
    \item One outer loop iteration for each $v \in P^{\xright}_u$,
    \item For each such $v$, an additional $D_v$ iterations.
\end{itemize}

Thus:
$$D_u = 1 + \sum_{v \in P^{\xright}_u} (D_v + 1) \le 1 + \sum_{v \in P^{\xright}_u} (6\Phi(v) - 8) = 6\Phi(u) - 9 + \left(10 - 8|P^{\xright}_u|\right),$$
which satisfies the desired bound.

\smallskip
\textit{Case 2:} Both $l$ and $r$ have potential at least two. Then:
\begin{itemize}
    \item One iteration for setting $\visit_{u,\xleft}$,
    \item For each $v \in P^{\xleft}_u$, an outer loop iteration and $D_v$ iterations in $l$,
    \item For every visiting tree in $l$, the algorithm sets $\visit_{u,\xright}$ and iterates over $P^{\xright}_u$, with $D_w$ iterations for each $w \in P^{\xright}_u$.
\end{itemize}

Hence:
\begin{align*}
D_u &= 1 + \sum_{v \in P^{\xleft}_u} (D_v + 1) + \Phi(l) \cdot \left(1 + \sum_{w \in P^{\xright}_u} (D_w + 1)\right) \\
&\le 1 + \sum_{v \in P^{\xleft}_u} (6\Phi(v) - 8) + \Phi(l) \cdot \left(1 + \sum_{w \in P^{\xright}_u} (6\Phi(w) - 8)\right) \\
&= 1 + 6 \Phi(l) - 8|P^{\xleft}_u| + \Phi(l) \cdot (1 + 6\Phi(r) - 8|P^{\xright}_u|) \\
&= 6\Phi(u) - 9 + \left(10 - 8|P^{\xleft}_u| + \Phi(l)(7 - 8|P^{\xright}_u|)\right).
\end{align*}

Again, the bracketed term is non-positive, completing the inductive step.

\smallskip
Therefore, in all cases we have $D_u \le 6\Phi(u) - 9$.
Since the total number of loop iterations over all free nodes is linear in $\Phi(r)$, the total time complexity is $\mathcal{O}(\Phi(r))$.
\end{proof}

\medskip
\textbf{Acknowledgment of Language Assistance}

The author acknowledges the use of a large language model (GPT-4) for assistance in improving the clarity, grammar, and overall readability of the English text.
All technical content, ideas, results, and conclusions presented in this work are the original contributions of the author.

\bibliographystyle{plain}
\bibliography{notes}

\appendix
\section{List of symbols}

\begin{description}
\item[$G^\star$] An input graph for which all perfect matching are enumerated.
\item[$N_V(u), N_E(u), N_V(W), N_E(W)$] Sets of vertices connected to $u$ and edges incident to a vertex $u$ and $N_V(W) = \bigcup_{u \in W} N_V(u)$ and $N_E(W) = \bigcup_{u \in W} N_E(u)$ where $W \subseteq V$, respectively.
\item[$G^T$] The graph obtained from $G$ by trimming it and the trimming function.
\item[$\M(G) = \M$] The set of all perfect matching of $G$.
\item[$\M_e^+, \M_e^-$] The set of all perfect matching containing and avoiding $e$.
\item[$\MM(G)$] The set of perfect matchings of $G^\star$ encoded in $G$ using the union-product circuit.
\item[$b^+(G), b^-(G), b(G)$] The set of all edges of $G$ contained in all perfect matchings of $G$, no perfect matching, and the union of $b^+(G)$ and $b^-(G)$.
\item[$M_\pi$] A perfect matching of $G$ containing $\pi$.
\item[$D(G,M)$] The directed graph on vertices $V$ where every matching edge of $M$ is oriented from one partite of $G$ to the other and non-matching edges $E \setminus M$ are oriented in the opposite direction.
\item[$\hG$] A directed acyclic graph obtained from $D(G^-_\pi, M_\pi)$ by contracting every component of $\K$ into a single vertex, and vertices $\sigma$ and $\tau$ are preserved, and parallel edges are removed.
Note that $\hG$ is also interpreted as a partially ordered set (poset) where a vertex $u$ of $\hG$ is smaller than a vertex $v$ if there exists an oriented path from $u$ to $v$ in $\hG$.
\item[$\M^+$-minimal edge $\pi$] An edge $\pi \in E$ is $\M^+$-minimal if there is no edge $e \in E$ such that $\M_e^+$ is a proper subset of $\M_\pi^+$.
\item[$\sigma \tau$] The edge $\pi$ oriented from $\sigma$ to $\tau$ in $D(G,M_\pi)$.
\item[$V_\sigma, V_\tau$] The partite of $V(G)$ containing $\sigma$ and $\tau$.
\item[$A, B$] Sets of edges of $G$ satisfying $\M = \M(G_A) \dotcup \M(G_B)$ and $\pi \in A \setminus B$.
\item[$G_A, G_B$] Graphs $G_A = G \setminus A$ and $G_B = G \setminus B$.
\item[$G^b,G^b_A, G^b_A$] Graph $G \setminus b^-(G)$ and $G_A \setminus b^-(G_A)$ and $G^b_B \setminus b^-(G_B)$
\item[$\Phi(G), \Phi(e), \Phi(u)$] The potential of a graph $G$ and an edge $e$ and a node $u$.
\item[$\Phi_E(G)$] The sum of potentials of edges in $E(G^b)$.
\item[$c = 0.1$] A constant used in the potential analysis: $\Phi(G_A) + \Phi(G_B) - \Phi(G) \ge c |E(G)|$.
\item[$G_e^-$] The graph $G \setminus \set{e}$.
\item[$G_e^+$] The graph obtained from $G$ by removing all edges incident to an edge $e$.
\item[$\K$] The set of all components of of strongly connected components of $D(G_\pi^+, M_\pi)$ and edges $b^+(G_\pi^+)$.
\item[$\tau, \kappa_1, \ldots, \kappa_\zeta, \sigma$] A maximal chain in $\hG$ containing all non-trivial components of $\K$.
\item[$\eta_e$] A node of the union-product circuit of an edge $e$.
\item[$X \Square Y$] The product of two sets $X$ and $Y$ is the set $\set{M_X \cup M_Y;\; M_X \in X,\, M_Y \in Y}$.
\item[$R_A, R_B, R'_A, R'_B$] Sets of external edges reachable in $G_A$ and $G_B$ and unreachable in $G_A$ and $G_B$.
\item[$\K_A, \K_B, \K'_A, \K'_B$] The set of components of $\K$ reachable in $G_A$ and $G_B$ and unreachable in $G_A$ and $G_B$.
\item[$K^\pi_B$] The component in $G^b_B$ containing $\pi$.
\item[$\delta(I)$] The set of edges of $G$ having only the outgoing vertex in a component inside an ideal $I$.
\item[$\Ex$] The set of all external edges; i.e. $\Ex = E(G) \setminus (E(\K) \cup \set\pi)$.
\item[$\Upsilon(u), \Upsilon(e)$] The set of matchings of $G^\star$ encoded in a node $u$, and the node $\eta_e$ associated to an edge $e$.
\item[$u \square v, u \cup v$] A new union and product node having $u$ and $v$ as children.
\end{description}

\end{document}

%% file: array.tex
\begin{tikzpicture}
    [vertex/.style={circle,draw,scale=0.5},
    inter/.style={fill,circle,scale=0.3}]

\node[vertex] (1) {};
\node[vertex,node distance=60mm] (2) [right=of 1] {};
\node[vertex] (3) [below=of 1] {};
\node[vertex,node distance=60mm] (4) [right=of 3] {};
\node[vertex] (5) [below=of 3] {};
\node[vertex,node distance=60mm] (6) [right=of 5] {};

\newcommand{\xxxx}[2] {
\draw[=,dotted] (#1) -- (#2) node[pos=0.2,inter] (#1#21) {} node[pos=0.4,inter] (#1#22) {} node[pos=0.6,inter] (#1#23) {} node[pos=0.8,inter] (#1#24) {};
\draw[=,very thick] (#1) -- (#1#21);
\draw[=,very thick] (#1#22) -- (#1#23);
\draw[=,very thick] (#1#24) -- (#2);
}
\xxxx{1}{2}
\xxxx{3}{4}
\xxxx{5}{6}

\newcommand{\yyyy}[2] {
\draw[=,dotted] (#1) -- (#2) node[pos=0.2,inter] (#1#21) {} node[pos=0.4,inter] (#1#22) {} node[pos=0.6,inter] (#1#23) {} node[pos=0.8,inter] (#1#24) {};
\draw[=,very thick] (#1#21) -- (#1#22);
\draw[=,very thick] (#1#23) -- (#1#24);
}
\yyyy{1}{4}
\yyyy{1}{6}
\yyyy{3}{2}
\yyyy{3}{6}
\yyyy{5}{2}
\yyyy{5}{4}

\draw[=,red] (3) -- (4) -- (5) -- (6) -- (3);

\end{tikzpicture}

%% file: trim.tex
\newcommand{\north}[1]{
\draw (#1.north) -- ++(0,0.3);
\draw (#1.north) -- ++(0.1,0.3);
\draw (#1.north) -- ++(-0.1,0.3);
}
\newcommand{\south}[1]{
\draw (#1.south) -- ++(0,-0.3);
\draw (#1.south) -- ++(0.1,-0.3);
\draw (#1.south) -- ++(-0.1,-0.3);
}

\begin{tikzpicture}
    [vertex/.style={circle,draw}]

\node[] (c) {};
\node[vertex] (u) [left=of c] {$u$};
\node[vertex] (v) [above=of c] {$v$};
\node[vertex] (w) [below=of c] {$w$};
\node[vertex] (v1) [right=of c] {$v_1$};

\node[vertex] (v1') [right=of v1] {$v'_1$};
\node[] (c') [right=of v1'] {};
\node[vertex] (u') [right=of c'] {$u'$};
\node[vertex] (v') [above=of c'] {$v'$};
\node[vertex] (w') [below=of c'] {$w'$};

\draw [=] (v1) -- (w) -- (u) -- (v) -- (v1) -- (v1') -- (w') -- (u') -- (v') -- (v1');

\north{v}
\south{w}
\north{v'}
\south{w'}
\south{v1'}

\node[] () [left=of v] {(a)};
\end{tikzpicture}

\begin{tikzpicture}
    [vertex/.style={circle,draw}]

\node[vertex] (z) {$z$};
\node[vertex] (v1) [right=of z] {$v_1$};
\node[vertex] (v1') [right=of v1] {$v'_1$};
\node[vertex] (z') [right=of v1'] {$z'$};

\draw [=] (v1) -- (v1');
\draw (z) edge[bend right,=] (v1);
\draw (v1) edge[bend right,=] (z);
\draw (z') edge[bend right,=] (v1');
\draw (v1') edge[bend right,=] (z');

\south{z}
\south{z'}
\north{z}
\north{z'}
\south{v1'}

\node[] () [left=of z] {(b)};
\end{tikzpicture}

\medskip

\begin{tikzpicture}
    [vertex/.style={circle,draw}]

\node[vertex] (z) {$z$};
\node[vertex] (v1) [right=of z] {$v_1$};
\node[vertex] (v1') [right=of v1] {$v'_1$};
\node[vertex] (z') [right=of v1'] {$z'$};

\draw [=] (z) -- (v1) -- (v1') -- (z');

\south{z}
\south{z'}
\north{z}
\north{z'}
\south{v1'}

\node[] () [left=of z] {(c)};
\end{tikzpicture}

\medskip

\begin{tikzpicture}
    [vertex/.style={circle,draw}]

\node[vertex] (bz) {$\bar z$};
\node[vertex] (z') [right=of bz] {$z'$};

\draw [=] (bz) -- (z');

\south{bz}
\north{bz}
\north{z'}
\south{z'}

\node[] () [left=of bz] {(d)};
\end{tikzpicture}

\medskip

\begin{tikzpicture}
    [vertex/.style={circle,draw},
    level distance=10mm,
    level 1/.style={sibling distance=50mm},
    level 2/.style={sibling distance=25mm},
    level 3/.style={sibling distance=12mm},
    ]

\node[vertex, label=below:$\eta_{\bar z z'}$] (root) {$\Square$}
    child { node[vertex, label=above:$\eta_{zv_1}$] {$\cup$}
        child { node[vertex] {$\Square$} 
            child { node[vertex] {$uv$}}
            child { node[vertex] {$wv_1$}}
        }
        child { node[vertex] {$\Square$} 
            child { node[vertex] {$vv_1$}}
            child { node[vertex] {$uw$}}
        }
    }
    child { node[vertex, label=above:$\eta_{z'v'_1}$] {$\cup$}
            child { node[vertex] {$\Square$} 
            child { node[vertex] {$u'v'$}}
            child { node[vertex] {$w'v'_1$}}
        }
        child { node[vertex] {$\Square$} 
            child { node[vertex] {$v'v'_1$}}
            child { node[vertex] {$u'w'$}}
        }
    };

\node[] () [left=of root] {(e)};
\end{tikzpicture}

%% file: circuit.tex
\begin{tikzpicture}
    [
    vertex/.style={circle,draw,thin,black,scale=0.5},
    big/.style={circle,draw,thin,black,node distance=6mm},
    ]

\node[vertex] (11) {};
\node[vertex,below left=of 11] (12) {};
\node[vertex,below right=of 11] (13) {};
\node[vertex,below=of 12] (14) {};
\node[vertex,below=of 13] (15) {};
\node[vertex,below left=of 15] (16) {};

\path (11) edge node[pos=0.3,left] {$a$} (12);
\path (11) edge node[pos=0.3,right] {$b$} (13);
\path (12) edge node[pos=0.3,left] {$c$} (14);
\path (12) edge node[pos=0.3,above] {$d$} (15);
\path (13) edge node[pos=0.3,above] {$e$} (14);
\path (13) edge node[pos=0.3,right] {$f$} (15);
\path (14) edge node[pos=0.7,left] {$g$} (16);
\path (15) edge node[pos=0.7,right] {$h$} (16);

\node[vertex,node distance=30mm,right=of 11] (21) {};
\node[vertex,below left=of 21] (22) {};
\node[vertex,below right=of 21] (23) {};
\node[vertex,below left=of 23] (24) {};

\path (21) edge node[pos=0.3,left] {$x$} (22);
\path (21) edge node[pos=0.3,right] {$y$} (23);
\path (22) edge node[pos=0.7,left] {$g$} (24);
\path (23) edge node[pos=0.7,right] {$h$} (24);

\node[big,node distance=40mm,right=of 16,yshift=-20mm] (e) {$e$};
\node[big,right=of e] (a) {$a$};
\node[big,right=of a] (f) {$f$};
\node[big,right=of f] (c) {$c$};
\node[big,right=of c] (b) {$b$};
\node[big,right=of b] (d) {$d$};

\node[big,above=of c] (bc) {$\Square$};
\draw[=] (b) -- (bc) -- (c);
\node[big,above=of b] (bd) {$\Square$};
\draw[=] (b) -- (bd) -- (d);
\node[big,above=of a] (ae) {$\Square$};
\draw[=] (a) -- (ae) -- (e);
\node[big,above=of f] (af) {$\Square$};
\draw[=] (a) -- (af) -- (f);

\node[big,above=of af] (cc) {$\cup$};
\draw[=] (bc) -- (cc) -- (ae);

\node[big,above=of bc] (ff) {$\cup$};
\draw[=] (af) -- (ff) -- (bd);

\node[big,above=of bd] (g) {$g$};
\node[big,above=of ff] (fff) {$\Square$};
\draw[=] (g) -- (fff) -- (ff);

\node[big,above=of ae] (h) {$h$};
\node[big,above=of cc] (ccc) {$\Square$};
\draw[=] (h) -- (ccc) -- (cc);

\node[big,above=of fff,xshift=-6mm] (root) {$\cup$};
\draw[=] (ccc) -- (root) -- (fff);

\node[] () [node distance=0mm,left=of cc] {$x$};
\node[] () [node distance=0mm,right=of ff] {$y$};

\node[] () [left=of 11] {(a)};
\node[] () [left=of 21] {(b)};
\node[] () [left=of root] {(c)};

\end{tikzpicture}

%% file: min.tex
\begin{tikzpicture}
    [vertex/.style={circle,draw}]

\node[vertex,red] (1) {1};
\node[vertex] (2) [right=of 1] {2};
\node[vertex,red] (3) [right=of 2] {3};
\node[vertex] (4) [right=of 3] {4};
\node[vertex,red] (5) [right=of 4] {5};
\node[vertex,blue] (6) [below=of 5] {6};
\node[vertex] (7) [left=of 6] {7};
\node[vertex,blue] (8) [left=of 7] {8};
\node[vertex] (9) [left=of 8] {9};
\node[vertex,blue] (10) [left=of 9] {10};

\draw [=] (1) -- (2) -- (3) -- (4) -- (5) -- (6) -- (7) -- (8) -- (9) -- (10) -- (1);
\draw [=,dotted] (1) -- (8) -- (5) -- (10) -- (3) -- (6) -- (1);
\draw [=,dotted] (3) -- (8);

\end{tikzpicture}

%% file: trim_alg.tex
\begin{tikzpicture}
    [vertex/.style={circle,draw,thin,black},
    norm/.style={edge from parent/.style={black,thin,draw}},
    emphr/.style={edge from parent/.style={red,very thick,draw}},
    emphg/.style={edge from parent/.style={green,very thick,draw}},
    level distance=10mm,
    sibling distance=10mm,
    level 1/.style={sibling distance=20mm},
    level 2/.style={sibling distance=15mm},
    ]

\node[vertex] (w) {$w$}
    child[emphg] { node[vertex] {$u$}
        child[norm] { node[vertex] {$v$} 
            child { node[vertex] {$v_1$}
                child[edge from parent/.style={blue,very thick,draw}] { node[vertex] {$y_1$}}
            }
            child { node[vertex] {$v_2$}
                child[emphr] { node[vertex] {$y_2$}
                    child[norm] { node[vertex] {}
                        child[emphr] { node[vertex] {}
                        }
                    }
                    child[norm] { node[vertex] {}
                        child[emphr] { node[vertex] {}
                        }
                    }
                }
            }
            child { node[vertex] {$v_3$}
                child[edge from parent/.style={yellow,very thick,draw}] { node[vertex] {$y_3$}}
            }
            child { node[vertex] {$v_4$}
                child[edge from parent/.style={pink,very thick,draw}] { node[vertex] {$y_4$}}
            }
        }
    }
    child { node[vertex] {$w_1$}
        child[emphg] { node[vertex] {}}
    }
    child { node[vertex] {$w_2$}
        child[emphg] { node[vertex] {}
            child[norm] { node[vertex] {}
                child[emphg] { node[vertex] {}
                }
            }
            child[norm] { node[vertex] {}
                child[emphg] { node[vertex] {}
                }
            }
        }
    };

\node[] () [left=of w] {(a)};
\end{tikzpicture}

\medskip

\begin{tikzpicture}
    [vertex/.style={rectangle,draw,thin,black},
     near/.style={sibling distance=10mm},
    level distance=10mm,
    level 1/.style={sibling distance=30mm},
    ]

\node[vertex] (p14) {$p^1_4$}
    child { node[vertex] (p13) {$p^1_3$}
        child { node[vertex] (p12) {$p^1_2$}
            child { node[vertex] (p11) {$p^1_1$}
                child { node[] (1) {$t_{uv}$} }
                child { node[] {$t_{vv_1}$} }
            }
            child { node {$t_{vv_2}$} }
        }
        child { node {$t_{vv_3}$} }
    }
    child { node {$t_{vv_4}$} };

\node[node distance=50mm,vertex,right=of p14] (p24) {$p^2_4$}
    child { node[vertex] (p23) {$p^2_3$}
        child { node[vertex] (p22) {$p^2_2$}
            child { node[vertex] (p21) {$p^2_1$}
                child { node[] (2) {$t_{vv_4}$} }
                child { node[] {$t_{vv_3}$} }
            }
            child { node {$t_{vv_2}$} }
        }
        child { node {$t_{vv_1}$} }
    }
    child { node {$t_{uv}$} };

\node[node distance=10mm,vertex,above=of p14,xshift=-10mm] (T3) {$t_{v_3y_3}$};
\draw[dotted,=] (p12) -- (T3) -- (2);
\node[node distance=10mm,vertex,left=of T3] (T2) {$t_{v_2y_2}$};
\draw[dotted,=] (p11) -- (T2) -- (p21);
\node[node distance=10mm,vertex,left=of T2] (T1) {$t_{v_1y_1}$};
\draw[dotted,=] (p22) -- (T1) -- (1);

\node[] () [left=of T1] {(b)};
\end{tikzpicture}